\documentclass[%
aip,
jmp,
 amsmath,amssymb,
reprint,%
]{revtex4-2}

\usepackage{graphicx}% Include figure files
\usepackage{dcolumn}% Align table columns on decimal point
\usepackage{bm}% bold math
\usepackage[utf8]{inputenc}
\usepackage[T1]{fontenc}
\usepackage{mathptmx}
\usepackage{color}
\usepackage{etoolbox}
\usepackage{braket}
\usepackage{amsmath}
\usepackage{dsfont}
\usepackage{hyperref}

\newcommand{\Ber}{\text{Ber}}
\newcommand{\U}{\text{U}}
\newcommand{\SU}{\text{SU}}
\newcommand{\tr}{\text{tr}\,}

\newcommand{\Pf}{{\text{Pf}}}
\newcommand{\abs}[1]{\ensuremath{\left\vert#1\right\vert}}

\makeatletter
\def\@email#1#2{%
 \endgroup
 \patchcmd{\titleblock@produce}
  {\frontmatter@RRAPformat}
  {\frontmatter@RRAPformat{\produce@RRAP{*#1\href{mailto:#2}{#2}}}\frontmatter@RRAPformat}
  {}{}
}%
\makeatother
\begin{document}

\preprint{AIP/123-QED}

\title[Winding Number Statistics for Chiral Random Matrices]{Winding Number Statistics for Chiral Random Matrices:\\
                     Averaging Ratios of Determinants with Parametric Dependence}

\author{Nico Hahn$^{1*}$, Mario Kieburg$^2$, Omri Gat$^3$ and Thomas Guhr$^1$}
\affiliation{$^1$ Fakult\"at f\"ur Physik, Universit\"at Duisburg--Essen, Duisburg, Germany\\
             $^2$ School of Mathematics and Statistics, The University of Melbourne, Melbourne, Australia\\
             $^3$ Racah Institute of Physics, The Hebrew University, Jerusalem, Israel
             }
\email{nico.hahn@uni-due.de}

\date{\today}% It is always \today, today,
             %  but any date may be explicitly specified

\begin{abstract}
  Topological invariance is a powerful concept in different branches
  of physics as they are particularly robust under perturbations. We generalize the ideas of computing the statistics of winding numbers for a specific parametric model of the chiral Gaussian Unitary Ensemble to other chiral 
  random matrix ensembles. Especially, we address the two chiral symmetry classes,
  unitary (AIII) and symplectic (CII), and we
  analytically compute ensemble averages for ratios of determinants
  with parametric dependence. To this end, we employ a technique that
 exhibits reminiscent supersymmetric structures while we never carry out any map to superspace.
\end{abstract}

\maketitle

\section{Introduction}
\label{secI}

The idea of classifying Hamilton operators that reveal spectral gaps through topological lenses has been very successful in physical systems as those classes are very robust with respect to perturbations. This robustness has been theoretically and experimentally verified in various systems, see, e.g., Refs.~\onlinecite{BH2011,OLZH2022,CFYRSB2022,CXYKT2021,BWPE2019}. One specific topological index is the winding number for chiral operators. It is indeed a winding number in the classical sense when considering the spectral flow of the complex eigenvalues of the off-diagonal block of the chiral Hamiltonian in the chiral representation with respect to the momentum/wavevector in the Brillouin zone. Due to periodicity and continuity of the eigenvalues as functions of the momentum and the condition of a spectral gap, the eigenvalues will move around the origin in closed contours.

Physically, a nonzero winding
number yields the number of localized modes at the boundaries and thus
indicates topologically nontrivial
systems~\cite{prodanBulkBoundaryInvariants2016,ChenChiou2020,
  Shapiro2020}. If disorder comes into play, the winding number can
become random and a statistical analysis is called for. We refer the
reader to Ref.~\onlinecite{BHWGG2021} for further discussion of the
physics aspects. Here, we consider simple schematic models of chiral
systems with a parametric dependence.  We are guided by the
long--standing experience that Random Matrix Theory is often capable
of modeling universal statistical properties~\cite{GuhrRMTReview,
  MehtaBook}. It is worthwhile mentioning that the winding number
statistics is not related to the parametric spectral correlations
introduced and investigated in Refs.~\onlinecite{SimonsAltshuler1993A,
  SimonsAltshuler1993B}.  Although the random matrix models are, apart
from chirality, very similar, the statistical observables are
different.

In a previous work~\cite{BHWGG2021}, three of the authors studied chiral unitary
symmetry and evaluated the winding number distribution as well as the
correlators of the winding number density. Here, we investigate two of the five chiral symmetry classes, which are among the ten symmetry
classes known as tenfold way~\cite{AltlandZirnbauer1997, Heinzner2005,
  Kitaev2009, Ryu2016}. More precisely, we work with the chiral  unitary (AIII) and symplectic (CII) symmetry.
Our objectives are ensemble averages for ratios of determinants with
parametric dependence. This is related to averages for ratios of
characteristic polynomials in the context of classical Random Matrix
Theory. Apart from the crucial importance of the latter in the
supersymmetry method~\cite{Efetov1983}, they are also interesting
quantities in their own right for mathematical physics, see the by far not exhaustive list of
Refs.~\onlinecite{FyodorovStrahov2003,BorodinStrahov2006,KieburgGuhr2010a,KieburgGuhr2010b,ASW2020,IF2018,Webb2015,MN2001,CFKRS2003,BHNY2007,Eberh2022,Fyod2004}.

To carry out our study, we employ and extend a method put forward some
years ago by two of the present
authors~\cite{KieburgGuhr2010a,KieburgGuhr2010b}. Jokingly, but not
deceptively, it has been coined ``supersymmetry without
supersymmetry'', because it uncovers supersymmetric structures deeply
rooted in the ensemble averages without actually mapping the integrals
to be considered to superspace. This method proceeds as follows. First, we
map the average for ratios of determinants with parametric dependence
to averages for ratios of characteristic polynomials over another
random matrix ensemble, referred to as spherical~\cite{Krishnapur2009, ForresterMays2012, Mays2013}.  Second, we reformulate the integrals
by introducing superspace Jacobians, also known as Berezinians, which are in the present case mixtures of Vandermonde and Cauchy determinants~\cite{BasorForrester1994}. This
facilitates a decomposition and direct formal computation of all
integrals, leading to determinants or Pfaffians. Third,  we exploit the results of Refs.~\onlinecite{KieburgGuhr2010a,KieburgGuhr2010b} where the kernels of these determinants and Pfaffians have been identified as averages for
ratios or products of only two determinants with parametric dependence. Finally, we evaluate these simplified averages over the spherical ensemble with the help of orthogonal and skew-orthogonal polynomials. Here, we show only the first and the last step, and refer to Refs.~\onlinecite{KieburgGuhr2010a,KieburgGuhr2010b} for the intermediate steps with general validity.

This paper is organized as follows: in Sec.~\ref{secII} we
mathematically define the random matrix problem to be solved.  We
summarize our results in Sec.~\ref{secIII}, while we give their derivation in
Sec.~\ref{secIV} and some of the details in the two appendices. In Sec.~\ref{secV} we summarize and conclude.

\section{Posing the Problem}
\label{secII}

We consider Hamiltonians in the classes AIII (chiral complex Hermitian, $\beta=2$) and CII (chiral quaternion Hermitian, $\beta=4$),
respectively, in the tenfold way \cite{Ryu2016, Kitaev2009,
  Schnyder2008, AltlandZirnbauer1997}. Those Hamiltonians satisfy a chiral symmetry
  \begin{equation} \label{2ChiralSymmetry}
\{\mathcal{C},H\} = 0\quad {\rm with}\quad \mathcal{C}^2=\mathds{1},
\end{equation}
where $\{,\}$ is the anticommutator and $\mathcal{C}$ is a chirality operator. There are actually three other symmetry classes of Hamiltonians with a chiral symmetry, which we aim to study in future surveys. One of those three, the BDI class (chiral real Hermitian, $\beta=1$),  can be dealt in the very same way though the joint probability density of the eigenvalues needed in our computations will be more involved. Thence, we deferred this discussion to a future publication. The index $\beta$ is the Dyson index indicating the real dimension of the chosen number field.

We employ and extend the conventions and notations of
Ref.~\onlinecite{BHWGG2021}. Importantly, all matrix elements in the
symplectic case CII are $2\times 2$ quaternions, effectively doubling
the dimension of $H$ and $\mathcal{C}$.

In a chiral basis, meaning an eigenbasis of the chirality operator $\mathcal{C}$ such that
\begin{equation}
\mathcal{C} = \begin{bmatrix}
\mathds{1} & 0
\\
0 & -\mathds{1}
\end{bmatrix},
\end{equation}
the Hamiltonian takes the block off--diagonal form
\begin{equation} \label{2ChiralHamiltonian}
H = \begin{bmatrix}
0 & K
\\
K^\dagger & 0
\end{bmatrix}.
\end{equation}
The matrix $K$ has the dimension $N\times N$ for AIII and $2N\times 2N$ for CII, and  $K^\dagger$ is 
its (Hermitian) adjoint. Hence,  the Hamiltonian is complex Hermitian ($\beta=2$) and quaternion self--dual ($\beta=4$), respectively.

A simple random matrix model of these Hamiltonians are given by chiral Gaussian Unitary and Symplectic Ensembles, labeled chGUE and chGSE; these matrices are drawn from Gaussian
probability distributions invariant under unitary or
unitary--symplectic rotations, cf., Eqs.~\eqref{prob.beta2} and~\eqref{prob.beta4}. Then, the matrices $K$ can also be viewed as forming the corresponding Ginibre ensembles~\cite{Ginibre1965}. We, however, are interested in a parametric dependence $K=K(p)$ and, thus, $H=H(p)$ to investigate topological properties.  The real variable $p$ parametrizes the one--dimensional unit circle $\mathcal{S}^1$, giving the interpretation of $H(p)$ as a Bloch Hamiltonian. Physically, the parameter $p$ is the momentum which is essentially given by a wavevector in the Brillouin zone. This interpretation has an important consequence for the class CII as the time reversal operator $\mathcal{T}$ acts on $K(p)$ like
\begin{equation}
\mathcal{T}K(p)\mathcal{T}^{-1}=[\tau_2\otimes\mathds{1}_N]K^*(p)[\tau_2\otimes\mathds{1}_N] = K(-p)
\end{equation}
with $(.)^*$ being the complex conjugation and $\tau_2\in\mathbb{C}^{2\times 2}$ being the second Pauli matrix. This has a crucial impact on the matrix entries of $K(p)$ because this matrix will not be real quaternion for a generic $e^{ip}\in\mathcal{S}^1$. Only for $p=0$, the symmetry directly implies a real quaternion structure for $K(0)$. Hence, for a general $e^{ip}\in\mathcal{S}^1$, we can expect that $K(p)$ is a complex $2N\times 2N$ matrix interpolating between real and imaginary quaternions.

The general question addressed in recent works~\cite{OLZH2022,CFYRSB2022,CXYKT2021,BWPE2019} is about the stability of the spectral properties of Hamiltonians under perturbations. In the present case, this is a question about the topology of subsets of chiral operators which can be quantified by the eigenvalues of the block matrix $K(p)$ which are also parametrically depending on $p$. In Ref.~\onlinecite{Shapiro2020} it has been proposed that  assuming a gaped Hamiltonian also  the eigenvalues of $K(p)$ exhibit a spectral gap to the origin. However, they are generically complex such that trajectories of the eigenvalues with respect to $e^{ip}\in\mathcal{S}^1$ describe paths around the origin without crossing it, due to the spectral gap. This is not only true for the class AIII but also for the other chiral symmetry classes. 

\begin{figure}[t!]
\centerline{\includegraphics[height=0.45\textwidth]{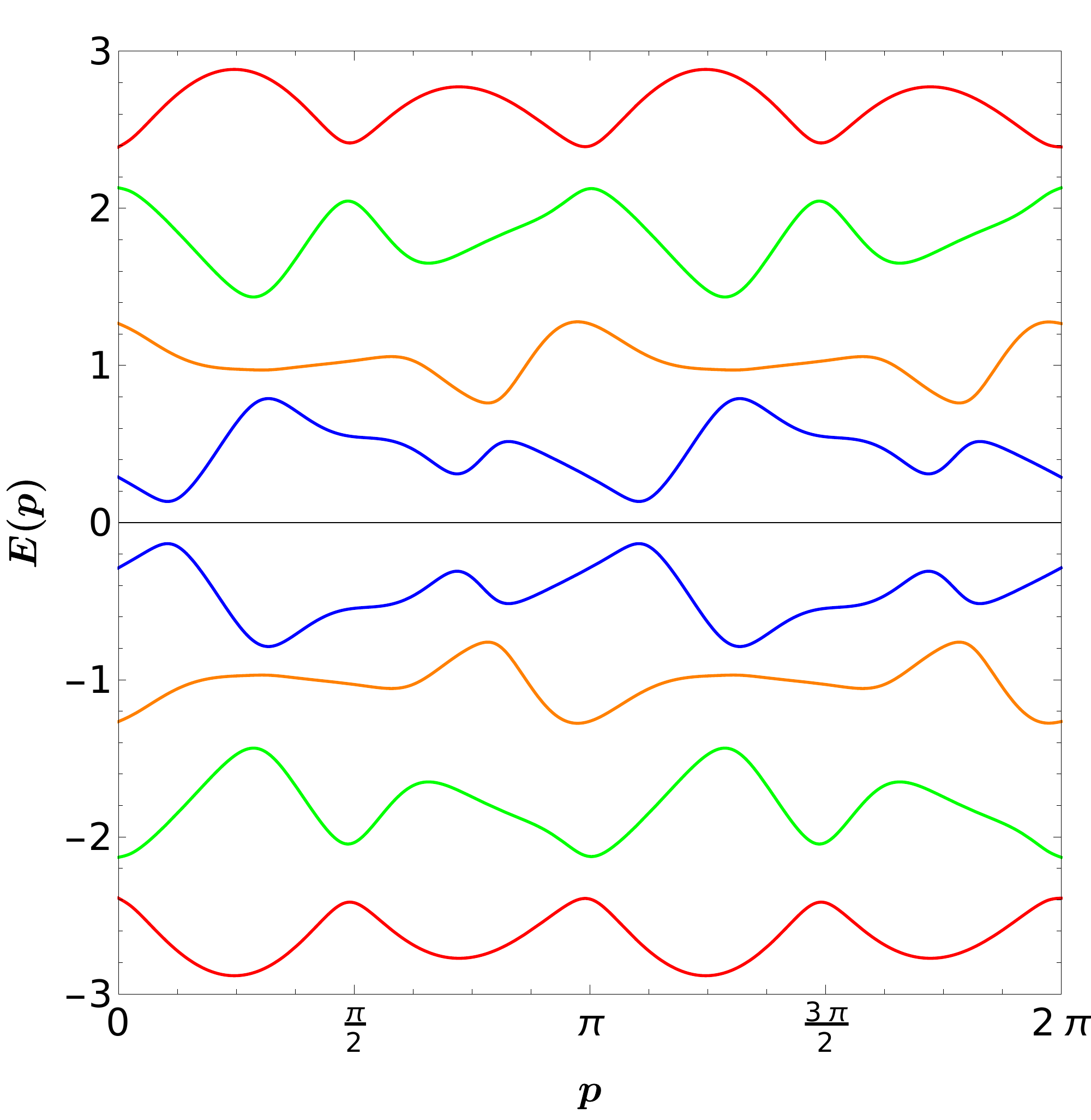}
\includegraphics[width=0.585\textwidth]{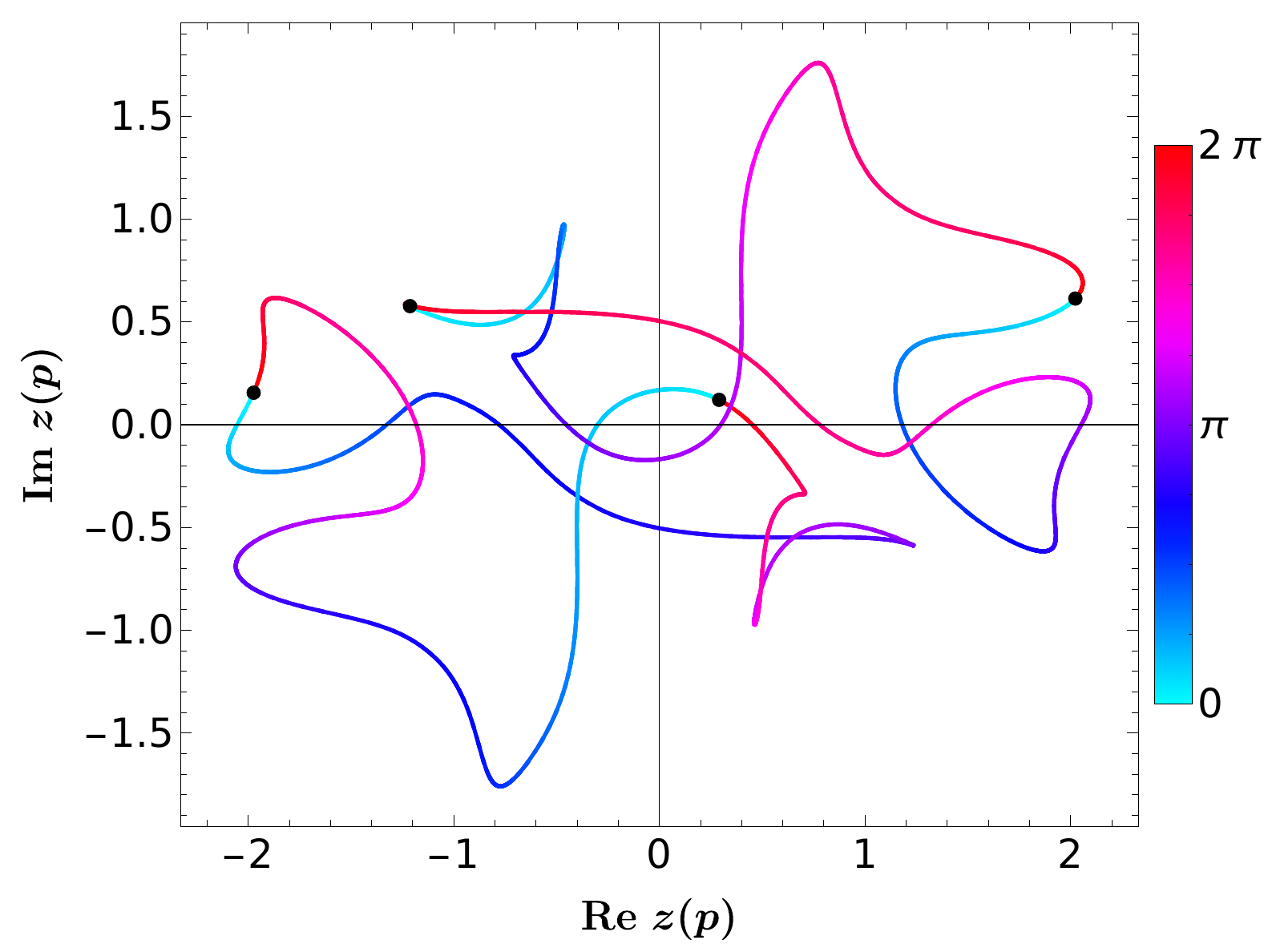}}
\includegraphics[height=0.5\textwidth]{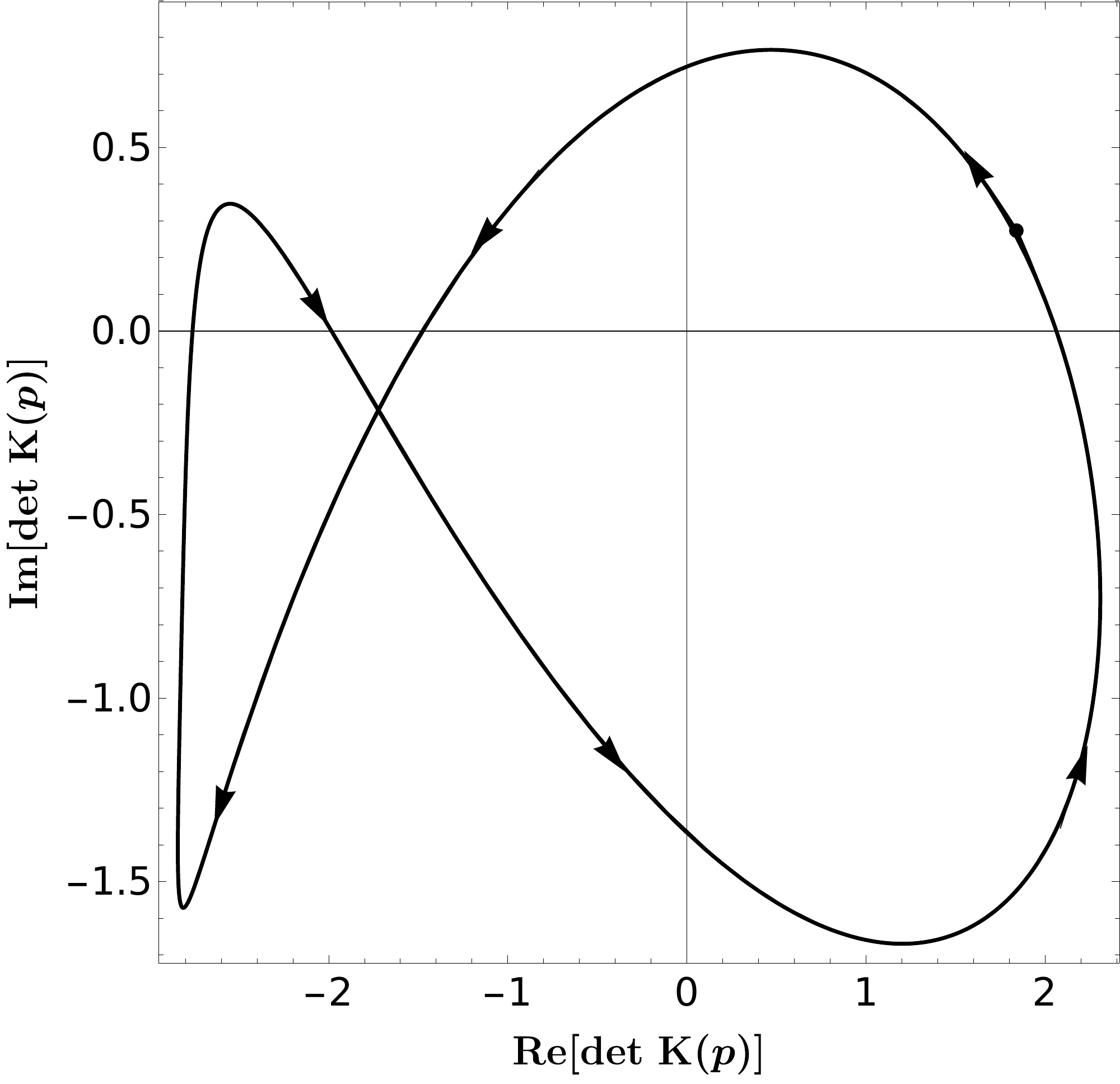}
\caption{A realization of an AIII Hamiltonian $H(p)=\cos(p)H_1+\sin(p)H_2$ with some fixed $4\times4$ complex matrices $K_1$ and $K_2$. The top left plot shows the real eigenvalues of $H(p)$, the top right one shows the generically complex eigenvalues of $K(p)=\cos(p)K_1+\sin(p)K_2$, and the bottom plot depicts the determinant $\det K(p)$. All plots show the parametric dependence in $p\in[0,2\pi]$ where we have employed the step size $2\pi/100$ and a B-Spline to obtain the curves. In both of the parametric plots the starting points $p=0$ are marked by black points and the directions are marked by a color gradient resp. arrows.}\label{fig:AIII}
\end{figure}

\begin{figure}[t!]
\centerline{
\includegraphics[height=0.45\textwidth]{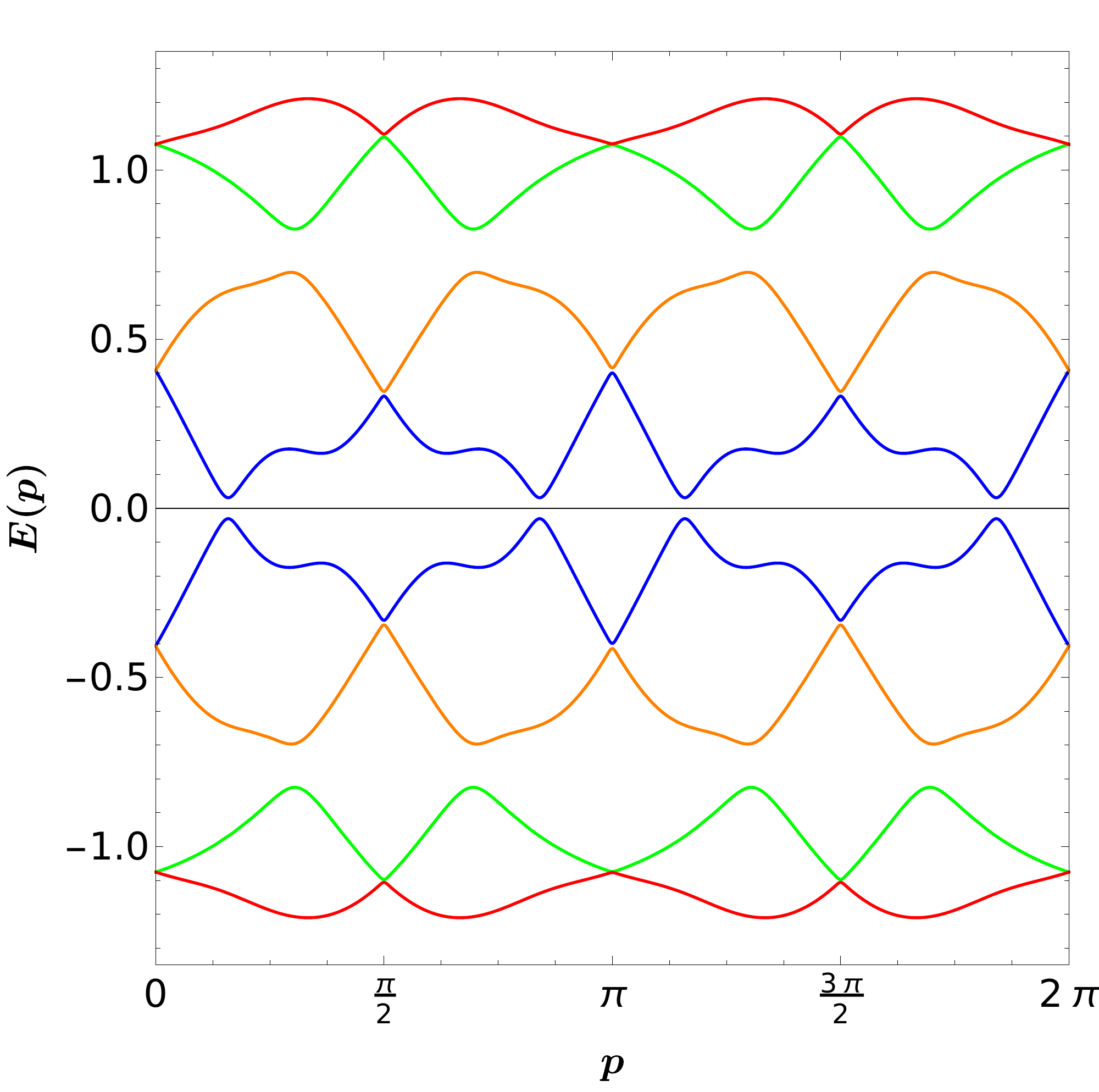}
\includegraphics[width=0.53\textwidth]{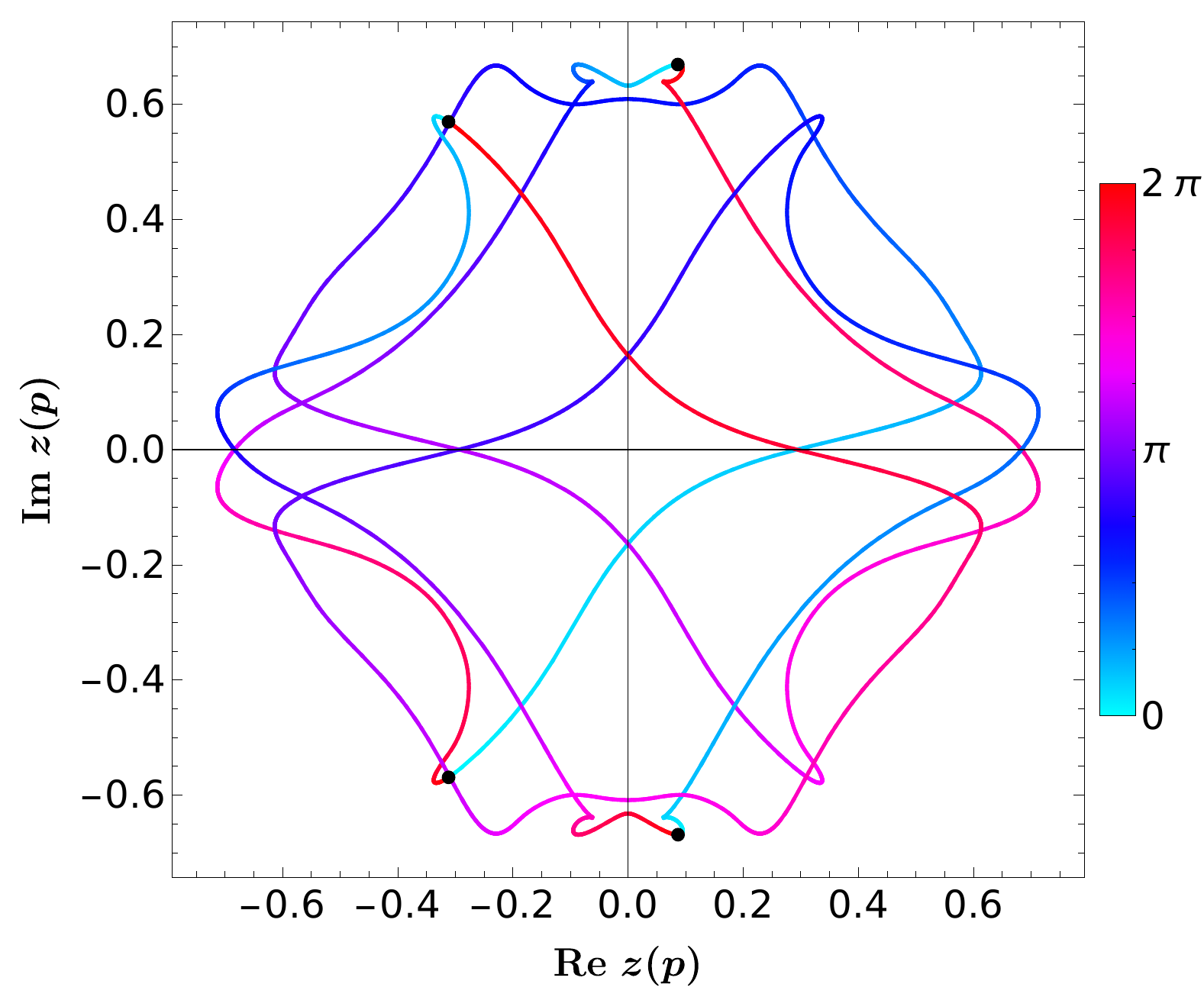}}
\includegraphics[height=0.495\textwidth]{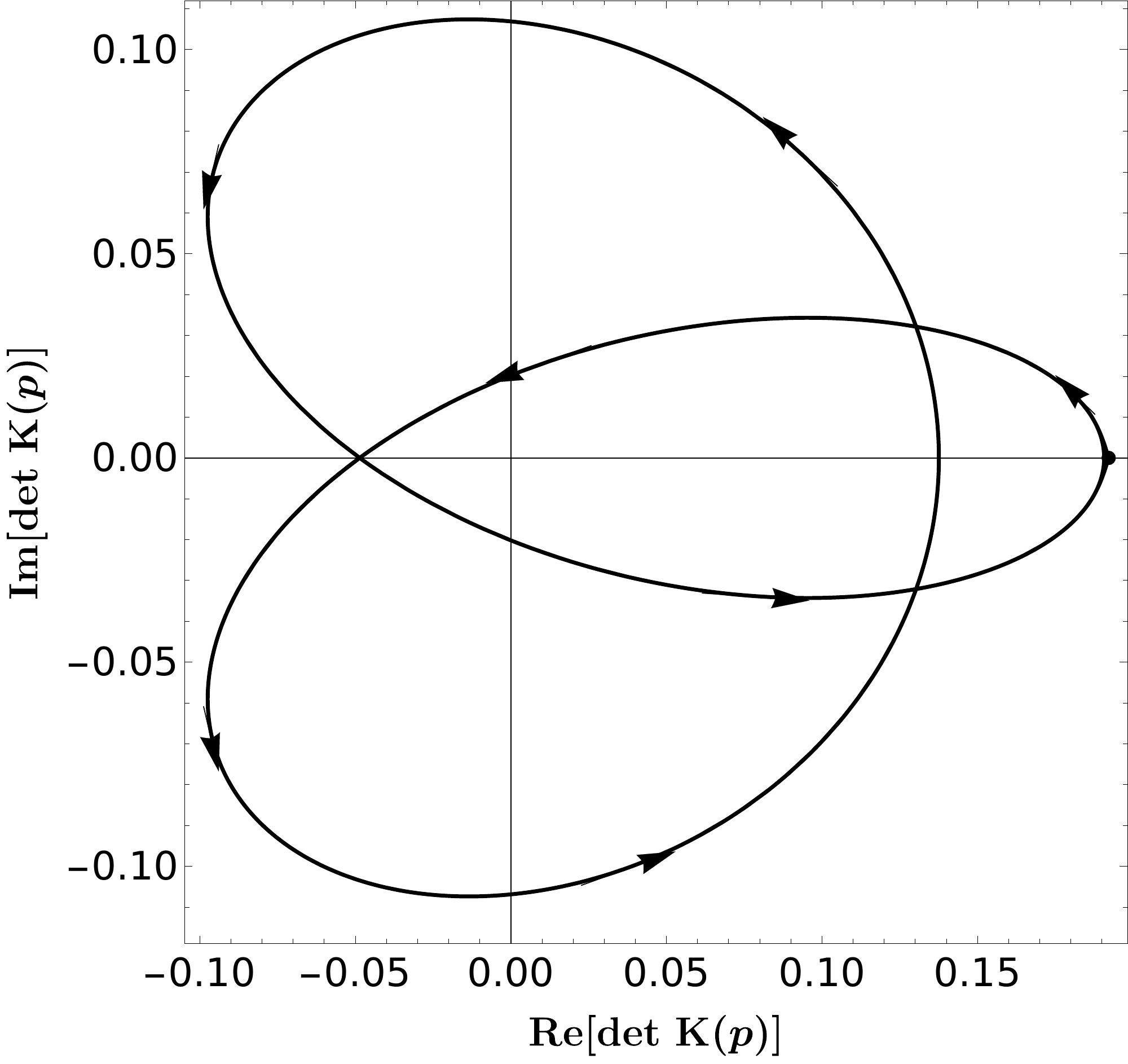}
\caption{A realization of a CII Hamiltonian $H(p)=\cos(p)H_1+i\sin(p)H_2$ with some fixed $4\times4$ real quaternion matrices $K_1$ and $K_2$. The top left plot shows the real eigenvalues of $H(p)$, the top right one shows the generically complex eigenvalues of $K(p)=\cos(p)K_1+i\sin(p)K_2$, and the bottom plot depicts the determinant $\det K(p)$. All plots show the parametric dependence in $p\in[0,2\pi]$ where we have employed the step size $2\pi/100$ and a B-Spline to obtain the curves. In both of the parametric plots the starting points $p=0$ are marked by black points and the directions are marked by a color gradient resp. arrows. There are exact crossings for the eigenvalues of $H(p)$ when either $\cos(p)=0$ or $\sin(p)=0$ as then the spectrum of $H(p)$ exhibits Kramers' degeneracy.}\label{fig:CII}
\end{figure}

In Figs.~\ref{fig:AIII} and~\ref{fig:CII} we illustrate the spectral flow for the matrix $K(p)=\cos(p)K_1+\sin(p) K_2$ with generic complex $K_1,K_2\in\mathbb{C}^{4\times4}$ (AIII) and for the matrix $K(p)=\cos(p)K_1+i\sin(p) K_2$ with generic real quaternion $K_1,K_2\in\mathbb{H}^{2\times2}\subset\mathbb{C}^{4\times 4}$ (CII). In these figures we illustrate the spectral flow of the eight eigenvalues of $H(p)$, the four complex eigenvalues of $K(p)$ and the determinant $\det K(p)$, which will play a crucial role when defining the winding number. Two important observations can be made which hold generically true. The order of the eigenvalues of $H(p)$ remains the same for all $p$ when level repulsion governs the spectral statistics, which implies that each eigenvalue of $H(p)$ is a $2\pi$ periodic function. In contrast, the eigenvalue spectrum of $K(p)$ may experience a permutation, meaning when running once from $p=0$ to $p=2\pi$ a chosen eigenvalue can become another one. Thence, the eigenvalues of $K(p)$ might have a different period than $2\pi$. Therefore, the eigenvalues of $K(p)$ are not suitable for classifying Hamiltonians. The determinant $\det K(p)$ is more suitable as this quantity must be $2\pi$ periodic. For the specific choice of the parametric dependence in Figs.~\ref{fig:AIII} and~\ref{fig:CII} we have $K(p+\pi) = -K(p)$, which restricts the amount of times $\det K(p)$ winds around the origin to be an even resp. odd number for even resp. odd matrix dimensions. These symmetries seen in the spectral flows are spurious and may not exist in general. 

The corresponding topological invariant describing this effect and classifying such subsets of chiral Hamiltonians
is the winding number, see Refs.~\onlinecite{Maffei2018, AsbothBook}
\begin{equation} \label{2WindingNumberDef}
W = \frac{1}{2\pi i} \int\limits_0^{2\pi} dp\ w(p),
\end{equation}
where the logarithmic derivative
\begin{equation} \label{2WindingNumberDensityDef}
w(p) = \frac{d}{dp} \ln \det K(p) = \frac{1}{\det K(p)} \frac{d}{dp} \det K(p)
\end{equation}
is the winding number density. Actually, the imaginary part of $w(p)$ is more closely related to the winding number since only that part describes the winding around the origin while the real part always integrates to zero because a closed path described by $\det K(p)$, which does not cross the origin, can be always continuously deformed into a path on the unit circle, implying $|\det K(p)|=1$ and, hence, a vanishing real part of the logarithmic derivative. This quantity is also reasonable for the quaternion (CII) and the real (BDI) case despite that determinants of real and real quaternion matrices are purely real. As mentioned above, the matrix $K(p)$ as a Bloch operator is generally complex.
 Obviously, the winding number $W$ can
directly be related to Cauchy's argument principle by writing the
integral as a contour integral for the complex variable $s=e^{ip}$,
see Ref.~\onlinecite{BHWGG2021}. Hence, the winding number is always an
integer, $W\in \mathbb{Z}$.

The parametric dependence of the random matrix $K(p)$ describes a random field on $\mathcal{S}^1$, which has its values in ${\rm Gl}_{\mathbb{C}}(N)$ for AIII or ${\rm Gl}_{\mathbb{C}}(2N)$ for CII. To have an analytically feasible model we assume a Gaussian random field that is centered. Thus, the model is fully controlled by its variance, which we assume to have the only non-vanishing covariances
\begin{equation}
\braket{K_{lj}^*(p)K_{lj}(q)}=S(p,q) \neq 0 \qquad	S(p,p) \geq 0
\end{equation}
with $p,q\in\mathcal{S}^1$ and any $l,j$, where $\braket{.}$ is the ensemble average. As this choice is independent of the matrix indices $l$ and $j$, $S(p,q)$ must be a scalar product on a vector space because of
\begin{equation}
\begin{split}
\braket{K_{lj}^*(p)[\lambda K_{lj}(q_1)+\mu K_{lj}(q_2)]}=&\lambda S(p,q_1)+\mu S(p,q_2),\\
 S^*(p,q)=\braket{K_{lj}^*(p)K_{lj}(q)}^*=&\braket{K_{lj}^*(q)K_{lj}(p)}=S(q,p)
 \end{split}
\end{equation}
for any $p,q,q_1,q_2\in\mathcal{S}^1$, $\mu,\lambda\in\mathbb{C}$, and $l,j$. Hitherto, we considered the most  general form for the covariance $S(p,q)$. The easiest non-trivial choice is a scalar product of a two-dimensional complex vector space, which can be realized by setting up random matrix fields as the linear combinations
\begin{equation} \label{2RMF}
K(p) = a(p) K_1 +b(p)  K_2
\end{equation}
with two scalar functions $a(p)$ and $b(p)$, that are smooth and $2\pi$-periodic. Arranging the two functions as a vector 
\begin{equation}	\label{2vdef}
v(p)=(a(p),b(p)) \in\mathbb{C}^2,
\end{equation}
the scalar product takes the form $S(p,q)=v^\dagger(p)v(q)$. Furthermore, when interpreting our random matrix model as a Bloch Hamiltonian (i.e. $p$ is a momentum), in the time reversal invariant cases the functions should satisfy
\begin{equation}
\mathcal{T}v(p)\mathcal{T}^{-1} = v^*(p)=v(-p)
\end{equation}
under conjugation with the anti-unitary time reversal operator $\mathcal{T}$.

The matrices $K_1$ and $K_2$ are either drawn from the complex Ginibre ensemble in the case AIII, see Eq.~\eqref{prob.beta2}, or from the real quaternion Ginibre ensemble in the case CII, see Eq.~\eqref{prob.beta4}, with probability density $P(K_1,K_2)$. As aforementioned, we denote the corresponding ensemble averages of an observable $F(K_1,K_2)$ with angular brackets,
\begin{equation} \label{ea}
\langle F \rangle = \int d[K_1,K_2] P(K_1,K_2)F(K_1,K_2),
\end{equation}
where the flat measures $d[K_1,K_2]$ are simply the products of the
differentials of all independent real variables. 

The structure of the random matrix field carries over from $K(p)$ to the
Hamiltonian $H(p)$ which becomes
\begin{equation} \label{2RMFHa}
  H(p) =a(p) H_1  +b(p) H_2 ,\qquad H_m = \begin{bmatrix}
0 & K_m
\\
K_m^\dag & 0
\end{bmatrix} \quad  m=1,2.
\end{equation}
This construction defines parametric combinations of two chGUE's (AIII) and chGSE's (CII),
respectively. 

Our goal is to calculate the ensemble averages for ratios of determinants with
parametric dependence
\begin{equation} \label{genfct}
Z^{(\beta,N)}_{k|l}(q,p) = \left\langle \frac{\prod_{j=1}^l \det K(p_j)}{\prod_{j=1}^k \det K(q_j)} \right\rangle
\end{equation}
for two sets of variables $p_1,\ldots,p_l$ and $q_1,\ldots,q_k$ in the
case $k=l$.  We introduce the more general definition~\eqref{genfct}
for $k$ and $l$ being different for reasons that will become clear in
the sequel. We notice that $k$ and $l$ are the numbers of determinants
in denominator and numerator, respectively. 

Ensemble averages for
ratios of the closely related characteristic polynomials are
mathematically the key objects in the supersymmetry
method~\cite{Efetov1983} since they serve as generators for
correlation functions of operator or matrix resolvents.  Similarly, we
can compute the $k$--point correlator
\begin{equation} \label{2kPointCorrelationDef}
C^{(\beta,N)}_k (p_1,\ldots,p_k) = \left\langle w(p_1) \cdots w(p_k) \right\rangle
\end{equation}  
of the winding number density as the $k$--fold derivative
\begin{equation} \label{2kPointCorrelationGen}
C^{(\beta,N)}_k(p_1,\ldots,p_k) = \frac{\partial^k}{\prod_{j=1}^k \partial p_j} Z^{(\beta,N)}_{k|k}(q,p)\Bigg|_{q=p}
\end{equation}
of the generator~\eqref{genfct}. However, as they are of
particular interest for the study of universality to be
undertaken in a forthcoming work, we relegate the results for the
correlation functions to a future publication.

Nevertheless, there is also
independent interest in ensemble averages for ratios of characteristic
polynomials~\cite{FyodorovStrahov2003,BorodinStrahov2006,KieburgGuhr2010a,KieburgGuhr2010b,ASW2020,IF2018,Webb2015,MN2001,CFKRS2003,BHNY2007,Eberh2022,Fyod2004}.
 in classical Random Matrix
Theory. For the Gaussian Orthogonal, Unitary and Symplectic
Ensemble a direct connection between averages corresponding to $k=l=1$
and the kernels of the $k$-point correlation functions was found in
Ref.~\onlinecite{GroenqvistGuhrKohler2004}, generalizing some implicit
observation~\cite{Guhr1991} for the unitary case in a supersymmetry
context. For classical Random Matrix Theory, the decomposition of
ensemble averages for ratios of characteristic polynomials in the case
of arbitrary $k$ and $l$ into ensemble averages for small $k$ and $l$
with $k+l=2$ was derived in Ref.~\onlinecite{BorodinStrahov2006},
employing a discrete approximation method related to representation
theory. In Refs.~\onlinecite{KieburgGuhr2010a,KieburgGuhr2010b}, two
of the present authors presented a very direct solution of this type
of problem. They extended a method put forward in
Ref.~\onlinecite{BasorForrester1994} by establishing a connection with
supersymmetry without mapping on superspace. More precisely, Jacobians
or Berezinians for the radial coordinates on certain symmetric
superspaces were identified in the integrals, considerably
facilitating the calculations. Here, we exploit the results of
Refs.~\onlinecite{KieburgGuhr2010a,KieburgGuhr2010b} to explicitly
compute the functions~\eqref{genfct}.

\section{Results}
\label{secIII}

Regardless of which of the two cases, AIII or CII, it is very useful to write the two coefficients $a(p)$ and $b(p)$ in terms of the 2-dimensional vector $v(p)$. Only then certain inherent symmetries are appropriately reflected in the results. For instance, in the unitary case AIII, labeled $\beta=2$, the partition function $Z^{(2,N)}_{k|k}(q,p)$, see Eq.~\eqref{genfct}, is invariant under the group $\SU(2)\times{\rm Gl}_{\mathbb{C}}(1)$. The part ${\rm Gl}_{\mathbb{C}}(1)$ corresponds to the invariance under rescaling $v(p)\to s v(p)$ for all $s\in{\rm Gl}_{\mathbb{C}}(1)=\mathbb{C}\setminus\{0\}$. The scaling factor drops out in the ratio of the characteristic polynomials. The subgroup $\SU(2)$ reflects an invariance when rotating $K_1$ and $K_2$ into each other. This carries over to an invariance for the vector $v(p)$, see Sec.~\ref{secIVB} for more details. Therefore, the result can only depend on the combinations $v^\dagger(p)v(q)$, $v^T(p)\tau_2v(q)$ and their complex conjugates. We emphasize that $v^T(p)\tau_2v(q)$ is also an invariant because $U=\tau_2U^*\tau_2$ for any $U\in{\rm SU}(2)$. Additionally, $Z^{(2,N)}_{k|l}(q,p)$ is a polynomial in $v(p_j)$ while it is quite likely to be not holomorphic in $v(q_j)$. In Sec.~\ref{secIVB}, we derive the result
\begin{equation} \label{genfctAIII}
Z^{(2,N)}_{k|k}(q,p) = \frac{\det \left[\displaystyle \frac{1}{v^T(q_m)\tau_2v(p_n)}\left(\frac{v^\dagger(q_m) v(p_n)}{v^\dagger(q_m) v(q_m)}\right)^N \right]_{1\leq m,n\leq k}}{\det \left[\displaystyle \frac{1}{v^T(q_m)\tau_2v(p_n)} \right]_{1\leq m,n\leq k}}
\end{equation}
for the unitary case. As often, the orthogonal and symplectic cases BDI and
CII, respectively, are considerably more demanding and lead to Pfaffian structures. The symplectic case, labeled $\beta=4$, is slightly simpler in its computation and its results. However, the biggest obstruction is that it respects the smaller invariance group ${\rm SO}(2)\times{\rm Gl}_{\mathbb{R}}(1)$. The ${\rm Gl}_{\mathbb{R}}(1)$ part is once more the simple rescaling of the two dimensional vector $v(p)\to s v(p)$ with $s\in{\rm Gl}_{\mathbb{R}}(1)=\mathbb{R}\setminus\{0\}$. Yet, the condition that the two matrices $K_1$ and $K_2$ must be real quaternion only allows a rotation of one matrix into the other one via the real special orthogonal group ${\rm SO}(2)$. Again more details of this symmetry discussion can be found in Sec.~\ref{secIVC}.

For the result we need a special kind of Lerch's transcendental function, see Ref.~\onlinecite{NIST},
\begin{equation}\label{Lerch}
\Phi_{n+1}^{(1)}(z)=-\frac{1}{z^{n+1}}\left[\ln(1-z)+\sum\limits_{j=1}^n\frac{z^j}{j}\right]
\end{equation}
as well as the polynomial
\begin{equation}\label{skewortho.beta4}
\begin{split}
q_{2n}^{(N)}(x) =&\sum_{m=0}^n\frac{{\rm B}(n+1,N-n+1/2)}{{\rm B}(m+1,N-m+1/2)}x^{2m}\\
=&\frac{2N+1}{2}{\rm B}\left(n+1,\frac{2N-2n+1}{2}\right)(1+x^2)^{n-1/2}\\
&-\frac{2N-2n-1}{2(n+1)}x^{2(n+1)}{ _2F_1}\left(1,\frac{3+2n-2N}{2};n+2;-x^2\right).
\end{split}
\end{equation}
The function ${\rm B}(x,y)=\Gamma(x)\Gamma(y)/\Gamma(x+y)$ is Euler's Beta function with the Gamma function $\Gamma(x)$. The polynomials are essentially truncated binomial series. The second representation involves Gauss' hypergeometric function ${ _2F_1}$. The polynomials are actually the skew-orthogonal polynomials of even order corresponding to the quaternion spherical ensemble, see Appendix~\ref{AppendixA} for their derivation. In Sec.~\ref{secIVC}, we derive the following result
\begin{align} \label{genfctCII}
Z^{(4,N)}_{k|k}(q,p) &= \frac{1}{\det \left[\displaystyle \frac{1}{iv^T(q_m)\tau_2v(p_n)} \right]_{1\leq m,n\leq k}}
\Pf \begin{bmatrix}
\widehat{\rm K}_1(p_m,p_n) & \widehat{\rm K}_2(p_m,q_n) \\
-\widehat{\rm K}_2(p_n,q_m) &  \widehat{\rm K}_3(q_m,q_n) \end{bmatrix}_{1\leq m,n\leq k},
\end{align}
where the kernel functions are given by
\begin{equation}
\begin{split}
\widehat{\rm K}_1(p_m,p_n)=& 2N(2N+1)[iv^T(p_n)\tau_2v(p_m)]^{2N-1}q_{2N-2}^{(N)}\left(\frac{v^T(p_m)v(p_n)}{iv^T(p_m)\tau_2v(p_n)}\right),
\\
\widehat{\rm K}_2(p_n,q_m) =& \frac{1}{iv^T(q_m)\tau_2v(p_n)} \left( \frac{v^T(p_n) v(p_n)}{i v^T(q_m) \tau_2 v(p_n)} \right)^{2N} \left( 1- \frac{v^T(q_m) v(p_n) v^\dag(q_m) v(p_n)}{v^T(q_m) \tau_2 v(p_n) v^\dag(q_m) \tau_2 v(p_n)} \right)^{-2N-1}
\\
&\times \left[ \left( \frac{v^\dag(q_m) v(p_n)}{i v^\dag(q_m) \tau_2 v(p_n)} \right)^{2N+1} \frac{v^T(q_m) v(p_n)}{i v^T(q_m) \tau_2 v(p_n)} + (2N+1) q_{2N}^{(N+1)}\left(\frac{v^\dagger(q_m)v(p_n)}{iv^\dagger(q_m)\tau_2v(p_n)}\right)\right],
\\
\widehat{\rm K}_3(q_m,q_n)=&-iv^T(q_m)\tau_2v(q_n)\left[\frac{v^\dagger(q_m)v^*(q_n)}{v^\dagger(q_m)v(q_m)v^\dagger(q_n)v(q_n)}\right]^{2N+2}\Phi_{2N+2}^{(1)}\left[\frac{|v^T(q_m)v(q_n)|^2}{v^\dagger(q_m)v(q_m)v^\dagger(q_n)v(q_n)}\right]\\
&+\left(\frac{iv^\dagger(q_n)\tau_2v^*(q_m)}{v^\dagger(q_m)v(q_m)v^\dagger(q_n)v(q_n)}\right)^{2N+1}q_{2N}^{(N+1)}\left(\frac{v^\dagger(q_n)v^*(q_m)}{iv^\dagger(q_n)\tau_2v^*(q_m)}\right).
\end{split}
\end{equation}
The block matrices in~\eqref{genfctCII} have to be read such that one takes a $k\times k$ matrix with $2\times 2$ matrices of the shown form as matrix entries.

\section{Derivations}
\label{secIV}

In Secs.~\ref{secIVB} and \ref{secIVC}, we first analyze the symmetries of the partition function~\eqref{genfct} for the symmetry classes AIII and CII. Those symmetries become handy when simplifying the computations. Furthermore, we trace the ensemble average over the two independent Ginibre matrices back to the spherical ensembles that have been studied in Refs.~\onlinecite{Krishnapur2009,Mays2013}. Using results from Refs.~\onlinecite{KieburgGuhr2010a,KieburgGuhr2010b}, we make use of determinantal and Pfaffian structures that reduce the problem of averaging over a ratio of $2k$ characteristic polynomials to averages of only two characteristic polynomials. In combination with the techniques of orthogonal and skew-orthogonal polynomials as well as some Complex Analysis tools we find the results summarized in Sec.~\ref{secIII}.

\subsection{Unitary Case (AIII)}
\label{secIVB}

When the two matrices $K_1,K_2\in{\rm Gl}_{\mathbb{C}}(N)$ are independently drawn from a complex Ginibre ensemble, i.e., their joint probability distribution is
\begin{equation}\label{prob.beta2}
P(K_1,K_2)=\pi^{-2N^2}\exp[-\tr K_1^\dagger K_1-\tr K_2^\dagger K_2],
\end{equation}
it is useful to write the two complex functions $a(p),b(p)$ in terms of the two-dimensional complex vector $v(p)$, see Eq.~\ref{2vdef}. The reason is that this ensemble actually satisfies an $\SU(2)$ symmetry given by
\begin{equation}
\hat{K}=\left[\begin{array}{c} K_1 \\ K_2 \end{array}\right]\longrightarrow [U\otimes\mathds{1}_N]\left[\begin{array}{c} K_1 \\ K_2 \end{array}\right]
\end{equation}
with $U\in\SU(2)$ acting on the two components of the matrix valued vector $\hat{K}$. One can readily verify $P(\hat{K})=P([U\otimes\mathds{1}_N]\hat{K})$ for any $U\in\SU(2)$. This will become handy when computing the partition function $Z^{(2,N)}_{k|k}(q,p)$ and recognizing that
\begin{equation}
K(p)= a(p)K_1 +  b(p) K_2=v^T(p)\hat{K}.
\end{equation}
Surely this $\SU(2)$ invariance will carry over to the vectors $v(p_j)$ and $v(q_j)$.

Before we exploit this symmetry we would like to draw attention to the relation of this ensemble to the complex spherical ensemble for which we need to rephrase the matrix $K(p)$ as follows
\begin{equation} \label{repar}
K(p)=a(p)K_1 + b(p)K_2 = b(p)K_1\left(\kappa(p)\mathds{1}_N+K_1^{-1}K_2\right),\ {\rm with}\ \kappa(p)=\frac{a(p)}{b(p)}.
\end{equation}
This way of writing is only possible when $b(p)\neq0$. This is, however, not very restrictive as the limit $b(p)\to0$ can be readily carried out in the results. The partition function~\eqref{genfct} for $k=l$, has then the form
\begin{equation} \label{genfctrepar.beta2}
Z^{(2,N)}_{k|k}(q,p) =  \left( \prod_{j=1}^k \frac{b(p_j)}{b(q_j)} \right)^{N} \left\langle \prod_{j=1}^k \frac{\det(\kappa(p_j)\mathds{1}_N+K_1^{-1}K_2)}{\det(\kappa(q_j)\mathds{1}_N+K_1^{-1}K_2)} \right\rangle.
\end{equation}
The random matrix $Y=K_1^{-1}K_2$ describes the complex spherical ensemble and it has been analyzed in several works~\cite{Krishnapur2009}. The corresponding probability density is
\begin{equation} \label{spherical.beta2}
\widetilde{G}^{(2)}(Y) = \pi^{-N^2} \prod_{j=0}^{N-1} \frac{(N+j)!}{j!}\ \frac{1}{\det^{2N}\left( \mathds{1}_{2N} + YY^\dagger \right)}
\end{equation}
and the corresponding joint probability distribution of the $N$ complex eigenvalues $(z_1,\ldots,z_{N})\in[\mathbb{C}\setminus\{0\}]^N$ is
\begin{align} \label{sphericalev.beta2}
G^{(2)}(z)  = \frac{1}{c^{(2)}} \frac{\abs{\Delta_N(z)}^2}{\prod_{j=1}^N(1+|z_j|^2)^{N+1}}\quad {\rm with}\quad c^{(2)}=\pi^N N! \prod_{j=1}^N {\rm B}(j,N+1-j).
\end{align}
As mentioned before, ${\rm B}(x,y)$ is Euler's Beta function.

An important remark about the integrability of the partition function is in order.  We certainly make use of the fact that a simple pole like $1/(\kappa(q_j)+z)$ is integrable in two dimensions such as the complex plane. However, we need to assume that all $\kappa(q_j)$ are pairwise distinct. In spite of this, it is rather remarkable that the final result can be nonetheless analytically continued to these singular points without any problems.

It is the structure of the joint probability density~\eqref{spherical.beta2}, which tells us that this ensemble follows a determinantal point process, see Ref.~\onlinecite{Borodin2015}, in particular, that the $k$-point correlation function is a $k\times k$ determinant with a single kernel function. This structure actually applies to the partition  function~\eqref{genfctrepar.beta2} as well. In Refs.~\onlinecite{BorodinStrahov2006,KieburgGuhr2010a} it was shown for more general ensembles than the one we study that
\begin{equation}\label{det.bet2}
\begin{split}
Z^{(2,N)}_{k|k}(q,p) =&  \left( \prod_{j=1}^k \frac{b(p_j)}{b(q_j)} \right)^{N}\ \frac{\displaystyle\det\left[ \left( \frac{b(q_m)}{b(p_n)} \right)^{N}\frac{Z^{(2,N)}_{1|1}(q_m,p_n)}{\kappa(q_m) - \kappa(p_n)} \right]_{1\leq m,n\leq k}}{\displaystyle\det\left[ \frac{1}{\kappa(q_m) - \kappa(p_n)}\right]_{1\leq m,n\leq k}}\\
=&\frac{\displaystyle\det\left[ \frac{Z^{(2,N)}_{1|1}(q_m,p_n)}{a(q_m)b(p_n) - b(q_m)a(p_n)} \right]_{1\leq m,n\leq k}}{\displaystyle\det\left[ \frac{1}{a(q_m)b(p_n) - b(q_m)a(p_n)}\right]_{1\leq m,n\leq k}}.
\end{split}
\end{equation}
The normalization can be checked by the asymptotic behavior
\begin{equation}
\lim_{a(p),a(q)\to\infty}\left(\prod_{j=1}^k \frac{a(q_j)}{a(p_j)}\right)^NZ^{(2,N)}_{k|k}(q,p)=1.
\end{equation}
The denominator in the first line of~\eqref{det.bet2} is known as the Cauchy determinant, see Ref.~\onlinecite{BasorForrester1994}, and can be identified with a Berezinian, see Ref.~\onlinecite{KieburgGuhr2010a} where this has been pointed out,
\begin{equation}
\sqrt{\Ber^{(2)}_{k|k}(\kappa(q);\kappa(p))} = \det\left[ \frac{1}{\kappa(q_m) - \kappa(p_n)}\right]_{1\leq m,n\leq k},
\end{equation}
which highlights the intimate link to a supersymmetric formulation of the problem. In the present work we will not go deeper into the details of this relation and defer it to future work when studying the universality of the large $N$ asymptotic.

The advantage of the determinantal form~\eqref{det.bet2} is that we actually need to compute the partition function for $k=1$. For this purpose, we finally make use of the $\SU(2)$ symmetry we have mentioned previously. The partition function
\begin{equation} 
Z^{(2,N)}_{1|1}(q_m,p_n)= F(v(q_m),v(p_n))
\end{equation}
can be understood as a function of the two vectors $v(q_m)$ and $v(p_n)$ and the $\SU(2)$ symmetry tells us that $F(v(q_m),v(p_n))=F(U^Tv(q_m),U^Tv(p_n))$ for all $U\in\SU(2)$. Therefore, we can choose the unitary matrix
\begin{equation}
U=\frac{1}{\sqrt{|a(q_m)|^2+|b(q_m)|^2}}\left[\begin{array}{cc} a^*(q_m) & -b(q_m) \\ b^*(q_m) & a(q_m) \end{array}\right]\in{\rm SU}(2)
\end{equation}
 such that the partition function simplifies to
\begin{equation} 
Z^{(2,N)}_{1|1}(q_m,p_n)=  \left\langle \det\left(\frac{v^\dagger(q_m)v(p_n)}{v^\dagger(q_m)v(q_m)}\mathds{1}_N+\tilde{b}K_1^{-1}K_2)\right) \right\rangle=  \left\langle \det\left(\frac{v^\dagger(q_m)v(p_n)}{v^\dagger(q_m)v(q_m)}\mathds{1}_N+\tilde{b}Y)\right) \right\rangle.
\end{equation}
The coefficient $\tilde{b}=iv^T(q_m)\tau_2v(p_n)/v^\dagger(q_m)v(q_m)\in\mathbb{C}$ is not very important as the $\U(1)$ invariance $Y\to e^{i\varphi} Y$ of the probability density tells us that the average of the characteristic polynomial $\det(x\mathds{1}_N-Y)$ only yields the monomial $x^N$. Thus, the final result is
\begin{equation}
\begin{split}
Z^{(2,N)}_{k|k}(q,p) =\frac{\displaystyle\det\left[ \frac{1}{a(q_m)b(p_n) - b(q_m)a(p_n)} \left(\frac{a^*(q_m)a(p_n)+b^*(q_m)b(p_n)}{|a(q_m)|^2+|b(q_m)|^2}\right)^N\right]_{1\leq m,n\leq k}}{\displaystyle\det\left[ \frac{1}{a(q_m)b(p_n) - b(q_m)a(p_n)}\right]_{1\leq m,n\leq k}}.
\end{split}
\end{equation}
This result actually nicely reflects the $\SU(2)$ symmetry as it only depends on the $\SU(2)$ invariants
$v^\dagger(q)v(q)$, $v^\dagger(q)v(p)$, and $v^T(q)\tau_2 v(p)=i(a(p)b(q)-a(q)b(p))$ with $\tau_2$ being the second Pauli matrix.

The $\SU(2)$ invariance is actually also reflected in the symmetry of the eigenvalue spectrum of the complex  spherical ensemble. In Ref.~\onlinecite{Krishnapur2009} it was pointed out that the complex spectrum is uniformly distributed on a two-dimensional sphere after a stereographic projection. It is the adjoint representation of $\SU(2)$, which is the special orthogonal group ${\rm SO}(3)$ that highlights the uniform distribution as it is the invariance group of a two-dimensional sphere.

\subsection{Symplectic Case (CII)}
\label{secIVC}

In the symplectic case we cannot exploit an $\SU(2)$ invariance. Due to the reality constraint of the real quaternion invertible matrices $K_1,K_2\in{\rm Gl}_{\mathbb{H}}(2N)$ in the form 
\begin{equation}
K_j=[\tau_2\otimes\mathds{1}_N]K_j^*[\tau_2\otimes\mathds{1}_N],
\end{equation}
we can only make use of the smaller invariance group
\begin{equation}
\hat{K}=\left[\begin{array}{c} K_1 \\ K_2 \end{array}\right]\longrightarrow [U\otimes\mathds{1}_{2N}]\left[\begin{array}{c} K_1 \\ K_2 \end{array}\right]\quad {\rm with}\quad U\in{\rm SO}(2).
\end{equation}
The probability density of two independent quaternion Ginibre ensembles, i.e.
\begin{equation}\label{prob.beta4}
P(K_1,K_2)=\pi^{-4N^2}\exp\left[-\frac{1}{2}\tr K_1^\dagger K_1-\frac{1}{2}\tr K_2^\dagger K_2\right],
\end{equation}
respects this symmetry. Thence, it will have some impact in our computations and will be visible in our results.

As before we express the expectation value over the two quaternion matrices $K_1,K_2\in{\rm Gl}_{\mathbb{H}}(2N)$ as an expectation value over the random matrix $Y=K_1^{-1}K_2\in{\rm Gl}_{\mathbb{H}}(2N)$, namely
\begin{equation} \label{genfctrepar.beta4}
\begin{split}
Z^{(4,N)}_{k|k}(q,p) %=&  \left( \prod_{j=1}^k \frac{b(p_j)}{b(q_j)} \right)^{2N} \left\langle \prod_{j=1}^k \frac{\det(\kappa(p_j)\mathds{1}_{2N}+K_1^{-1}K_2)}{\det(\kappa(q_j)\mathds{1}_{2N}+K_1^{-1}K_2)} \right\rangle\\
=&  \left( \prod_{j=1}^k \frac{b(p_j)}{b(q_j)} \right)^{2N} \left\langle \prod_{j=1}^k \frac{\det(\kappa(p_j)\mathds{1}_{2N}+Y)}{\det(\kappa(q_j)\mathds{1}_{2N}+Y)} \right\rangle,
\end{split}
\end{equation}
with $\kappa(p)=a(p)/b(p)$ defined as before.
The matrix $Y$ is now drawn from the quaternion spherical ensemble following the probability density~\cite{Mays2013}
\begin{equation} \label{spherical.beta4}
\widetilde{G}^{(4)}(Y) = \pi^{-2 N^2} \prod_{j=1}^N \frac{\left(2N+2j-1\right)!}{\left(2j-1 \right)!}\ \frac{1}{\det^{2 N}\left( \mathds{1}_{2N} + YY^\dag \right)}.
\end{equation}
Due to being quaternion each eigenvalue $z$ of $Y$ has a complex conjugate $z^*$, which is also an eigenvalue. The corresponding joint probability density of the eigenvalues $z={\rm diag}(z_1,z_1^*,z_2,z_2^*,\ldots,z_N,z_N^*)$ is given by
\begin{align} \label{sphericalev.beta4}
G^{(4)}(z)  = \frac{1}{c^{(4)}} \Delta_{2N}(z) \prod_{j=1}^N \frac{z_j-z_j^*}{(1+|z_j|^2)^{2N+2}}\quad{\rm with}\quad c^{(4)}=(2\pi)^NN!\prod_{j=1}^N{\rm B}(2j,2N+2-2j).
\end{align}
Considering this explicit form, the question of integrability for the considered partition function can be raised anew. It is this time not evident even in the case of pairwise distinct complex pairs $(\kappa(q_j),\kappa^*(q_j))$ as we encounter terms of the form $1/[(\kappa(q_j)+z_j)(\kappa(q_j)+z_j^*)]$. As long as $\kappa(q_j)$ is not real, the singularities are simple poles. However, when $\kappa(q_j)$ is real this term becomes a double pole of the integrand, which is, in general, not integrable even in two dimensions.  The fortunate fact that renders also this kind of pole integrable is the factor $|z_j-z_j^*|^2$ as it vanishes like a square when $z_j$ becomes real. Therefore, the combination $|z_j-z_j^*|^2/[(\kappa(q_j)+z_j)(\kappa(q_j)+z_j^*)]$ is absolutely integrable even when $\kappa(q_j)$ becomes real. The condition of pairwise distinct complex pairs  $(\kappa(q_j),\kappa^*(q_j))$ can be anew dropped for the final result where the limit $\kappa(q_a)\to\kappa(q_b)$ as well as $\kappa(q_a)\to\kappa^*(q_b)$ is well-defined, see the summary of the results in Sec.~\ref{secIII}.

It is well known, see Ref.~\onlinecite{Mays2013}, that the quaternion spherical ensemble describes a Pfaffian point process, and as before, this structure carries over to the partition function, which becomes, see Refs.~\onlinecite{BorodinStrahov2006,KieburgGuhr2010b},
\begin{align}\label{Pfaffian.beta4}
Z^{(4,N)}_{k|k}(q,p) &= \frac{1}{\displaystyle\det\left[ \frac{1}{\kappa(q_m) - \kappa(p_n)}\right]_{1\leq m,n\leq k}} \Pf \begin{bmatrix}
{\rm K}_1^{(4)}(p_m,p_n) & {\rm K}_2^{(4)}(p_m,q_n) \\
-{\rm K}_2^{(4)}(p_n,q_m) &  {\rm K}_3^{(4)}(q_m,q_n) \end{bmatrix}_{1\leq m,n\leq k},
\end{align}
where the three kernel functions are
\begin{equation}\label{kernel.beta4}
\begin{split}
{\rm K}_1^{(4)}(p_m,p_n)=&(\kappa(p_n)-\kappa(p_m))[b(p_m)b(p_n)]^{2N}\widetilde{Z}^{(4,N-1)}_{0|2}(p_m,p_n),\\
{\rm K}_2^{(4)}(p_n,q_m)=&\frac{1}{\kappa(q_m)-\kappa(p_n)}Z^{(4,N)}_{1|1}(p_n,q_m),\\
{\rm K}_3^{(4)}(q_m,q_n)=&\frac{\kappa(q_n)-\kappa(q_m)}{[b(q_m)b(q_n)]^{2N}}\widetilde{Z}^{(4,N+1)}_{2|0}(q_m,q_n).
\end{split}
\end{equation}
The Pfaffian is normalized such that
\begin{equation}
\Pf [i\tau_2,i\tau_2,\ldots,i\tau_2]=1,
\end{equation}
and we have employed the following definition for $l-k$ even and $M+(l-k)/2<N+1$
\begin{equation}
\begin{split}
\widetilde{Z}^{(4,M)}_{k|l}(q,p) =& \frac{1}{(2\pi)^{M+(l-k)/2}M!\prod_{j=1}^{M+(l-k)/2}{\rm B}(2j,2N+2-2j)}\\
&\times\int\limits_{\mathbb{C}^M} d[z] \Delta_{2M}(z) \prod_{r=1}^M \frac{z_r-z_r^*}{(1+|z_r|^2)^{2N+2}}\ \prod_{j=1}^M\frac{\prod_{n=1}^l (\kappa(p_n) + z_j) (\kappa(p_n) + z_j^*)}{\prod_{m=1}^k (\kappa(q_m) + z_j)(\kappa(q_m) + z_j^*)}.
\end{split}
\end{equation}
Let us highlight that the weight function $g^{(4)}(z)=(z-z^*)/(1+|z|^2)^{2N+2}$ remains always the same in this definition, while the number $M$ of integration variables varies.

The result~\eqref{Pfaffian.beta4} follows from Ref.~\onlinecite{KieburgGuhr2010b} when identifying in a distributional way the weight function $g^{(4)}(z)$ with the skew-symmetric two-point weight involving the Dirac delta function for complex numbers
\begin{equation}
\tilde{g}^{(4)}(z_1,z_2)=\frac{z_1-z_2}{(1+|z_1|^2)^{N+1}(1+|z_2|^2)^{N+1}}\delta(z_2-z_1^*).
\end{equation}
The integration over every second variable yields the joint probability density~\eqref{sphericalev.beta4}. In the ensuing three subsections we compute explicit expressions of these three kernels~\eqref{kernel.beta4}.

\subsubsection{The Kernel ${\rm K}_1^{(4)}$}\label{sec:K1.beta4}

The kernel function ${\rm K}_1^{(4)}(p_m,p_n)$ is expressed in terms of $\widetilde{Z}^{(4,N-1)}_{0|2}(\kappa(p_m),\kappa(p_n))$. We are in the lucky position that we can relate this function to the partition functions $Z^{(4,N-1)}_{0|2}(p_m,p_n)$ for which we can exploit the ${\rm SO}(2)$ symmetry.  This relation is given by
\begin{equation}
\begin{split}
\widetilde{Z}^{(4,N-1)}_{0|2}(p_m,p_n)=&\frac{1}{2\pi\ {\rm B}(2N,2)\bigl\langle \det K_1^2 \bigl\rangle}\frac{Z^{(4,N-1)}_{0|2}(p_m,p_n)}{[b(p_m)b(p_n)]^{2N-2}}\\
=&\frac{\bigl\langle \det(a(p_m)K_1+b(p_m)K_2)\det(a(p_n)K_1+b(p_n)K_2) \bigl\rangle}{2\pi\ {\rm B}(2N,2)\bigl\langle \det K_1^2 \bigl\rangle[b(p_m)b(p_n)]^{2N-2}} ,
\end{split}
\end{equation}
where we average over two independent invertible $(2N-2)\times(2N-2)$ real quaternion Ginibre matrices $K_1,K_2\in{\rm Gl}_{\mathbb{H}}(2N-2)$.  The limits
\begin{equation}
\lim_{\kappa(p)\to\infty}\frac{\widetilde{Z}^{(4,N-1)}_{0|2}(p_m,p_n)}{[\kappa(p_m)\kappa(p_n)]^{2N-2}}=\frac{1}{2\pi\ {\rm B}(2N,2)}\quad{\rm and}\quad \lim_{a(p)\to\infty}\frac{Z^{(4,N-1)}_{0|2}(p_m,p_n)}{[a(p_m)a(p_n)]^{2N-2}}=\bigl\langle \det K_1^2 \bigl\rangle
\end{equation}
relate the normalization of the two kinds of functions.

The partition function $Z^{(4,N-1)}_{0|2}(p_m,p_n)$ is a polynomial in the complex functions $a(p_m)$, $b(p_m)$, $a(p_n)$, and $b(p_n)$. Hence, we can also consider the average
\begin{equation}
\begin{split}
\Xi_1=\frac{\bigl\langle \det(a_1K_1+b_1K_2)\det(a_2K_1+b_2K_2) \bigl\rangle}{\bigl\langle \det K_1^2 \bigl\rangle}
\end{split}
\end{equation}
with only fixed real $a_1,b_1,a_2,b_2\in\mathbb{R}$ variables satisfying $b_1a_2-a_1b_2\neq0$ and then perform an analytic continuation in the result to the complex functions. We need this detour via analytic continuation because we can only rotate real vectors with the ${\rm SO}(2)$ symmetry similar to what we have done in the complex case AIII. Therefore, the average is equal to
\begin{equation}
\begin{split}
\Xi_1=&\frac{\bigl\langle \det([a_1a_2+b_1b_2]K_1+[b_1a_2-a_1b_2]K_2)\det(K_1) \bigl\rangle}{\bigl\langle \det K_1^2 \bigl\rangle}\\
=&\frac{[b_1a_2-a_1b_2]^{2N-2}}{(2\pi)^{N-1}(N-1)!\prod_{j=1}^{N-1}{\rm B}(2j,2N+2-2j)}\\
&\times\int\limits_{\mathbb{C}^{N-1}} d[z] \Delta_{2N-2}(z) \prod_{r=1}^{N-1} \frac{z_r-z_r^*}{(1+|z_r|^2)^{2N+2}}\ \prod_{j=1}^{N-1}\left(\frac{a_1a_2+b_1b_2}{b_1a_2-a_1b_2} + z_j\right)\left(\frac{a_1a_2+b_1b_2}{b_1a_2-a_1b_2} + z_j^*\right),
\end{split}
\end{equation}
where we have rotated with the special orthogonal matrix
\begin{equation}
U=\frac{1}{\sqrt{a_2^2+b_2^2}}\left[\begin{array}{cc} a_2 & -b_2 \\ b_2 & a_2 \end{array}\right]\in{\rm SO}(2).
\end{equation}
Apart from the factor $[b_1a_2-a_1b_2]^{2N-2}$ this integral is the Heine-like formula, see Ref.~\onlinecite{AKP2010} as well as Eq.~\eqref{Heinne.beta4}, for the monic skew-orthogonal polynomial $q_{2N-2}^{(N)}(x)$ of degree $2N-2$ corresponding to the weight function $g^{(4)}(z)=(z-z^*)/(1+|z|^2)^{2N+2}$. The skew-orthogonal polynomials have been computed in Appendix~\ref{AppendixA}.

Summarizing, the partition function $Z^{(4,N-1)}_{k|l}(p_m,p_n)$ has the form
\begin{equation}
\begin{split}
\frac{Z^{(4,N-1)}_{0|2}(p_m,p_n)}{\bigl\langle \det K_1^2 \bigl\rangle}=&[b(p_m)a(p_n)-a(p_m)b(p_n)]^{2N-2}q_{2N-2}^{(N)}\left(\frac{a(p_m)a(p_n)+b(p_m)b(p_n)}{a(p_m)b(p_n)-b(p_m)a(p_n)}\right)\\
=&\sum_{j=0}^{N-1}\frac{{\rm B}(N,3/2)}{{\rm B}(j+1,N-j+1/2)}[a(p_m)a(p_n)+b(p_m)b(p_n)]^{2j}\\
&\times[b(p_m)a(p_n)-a(p_m)b(p_n)]^{2N-2-2j}.
\end{split}
\end{equation}
We would like to underline that this formula is also true for the complex functions $a(p)$ and $b(p)$ despite we have derived it for real coefficients due to being a polynomial in these functions. The first kernel function is then
\begin{equation}\label{kernel.beta4.K1}
\begin{split}
{\rm K}_1^{(4)}(p_m,p_n)=&\frac{\kappa(p_n)-\kappa(p_m)}{2\pi {\rm B}(2N,2)}[b(p_m)b(p_n)]^{2}[b(p_m)a(p_n)-a(p_m)b(p_n)]^{2N-2}\\
&\times q_{2N-2}^{(N)}\left(\frac{a(p_m)a(p_n)+b(p_m)b(p_n)}{a(p_m)b(p_n)-b(p_m)a(p_n)}\right)\\
%=&\frac{b(p_m)b(p_n)}{2\pi}\sum_{j=0}^{N-1}\frac{{\rm B}(N,3/2)}{{\rm B}(2N,2) {\rm B}(j+1,N-j+1/2)}[a(p_m)a(p_n)+b(p_m)b(p_n)]^{2j}\\
%&\times[b(p_m)a(p_n)-a(p_m)b(p_n)]^{2N-1-2j}\\
=&\frac{b(p_m)b(p_n)}{2\sqrt{\pi}}\sum_{j=0}^{N-1}\frac{N!(1+2N)}{j!\Gamma(N-j+1/2)}[a(p_m)a(p_n)+b(p_m)b(p_n)]^{2j}\\
&\times[b(p_m)a(p_n)-a(p_m)b(p_n)]^{2N-1-2j}.
\end{split}
\end{equation}
This sum is apart from a prefactor a truncated binomial series.

\subsubsection{The Kernel ${\rm K}_2^{(4)}$}\label{sec:K2.beta4}

For the second kernel function we need to evaluate the partition function
\begin{equation}
\begin{split}
Z^{(4,N)}_{1|1}(\kappa(q_m),\kappa(p_n)) =&\left\langle\frac{\det( a(p_n)K_1 + b(p_n)K_2 )}{\det( a(q_m) K_1+ b(q_m)K_2 )} \right\rangle
\end{split}
\end{equation}
which is a polynomial in $a(p_n)$ and $b(p_n)$. With the very same arguments as in the previous subsection we can exploit the analyticity in these two variables and replace them by two fixed real variables $a_1,b_1\in\mathbb{R}$ and analytically continue the result at the end of the day. 
Unfortunately, we are not allowed to do the same trick for $a(q_n)$ and $b(q_n)$ as the partition function is not holomorphic in these two variables, actually the result will also depend on their complex conjugates such that we only replace them by two generic but fixed complex variables $a_2,b_2\in\mathbb{C}$.

We are allowed to apply an ${\rm SO}(2)$ rotation to simplify the average to
\begin{equation}
\begin{split}
\Xi_2 =&\left\langle\frac{\det( a_1K_1 + b_1K_2 )}{\det( a_2K_1+ b_2K_2 )} \right\rangle\\
=&[a_1^2+b_1^2]^{2N}\left\langle\frac{\det K_1}{\det( [a_2a_1+b_1b_2] K_1+ [b_1a_2-a_1b_2]K_2 )} \right\rangle\\
=&\frac{1 }{(2\pi)^{N}N!\prod_{j=1}^{N}{\rm B}(2j,2N+2-2j)}\left(\frac{a_1^2+b_1^2}{b_1a_2-a_1b_2}\right)^{2N}\int\limits_{\mathbb{C}^{N}} d[z] \Delta_{2N}(z) \prod_{r=1}^{N} \frac{z_r-z_r^*}{(1+|z_r|^2)^{2N+2}}\\
&\times \prod_{j=1}^{N}\left(\frac{a_1a_2+b_1b_2}{b_1a_2-a_1b_2}+ z_j\right)^{-1}\left(\frac{a_1a_2+b_1b_2}{b_1a_2-a_1b_2}+ z_j^*\right)^{-1}.
\end{split}
\end{equation}
We abbreviate the ratio
\begin{equation}
\kappa=\frac{a_1a_2+b_1b_2}{b_1a_2-a_1b_2}
\end{equation}
and identify another Berezinian, see Ref.~\onlinecite{KieburgGuhr2010a},
\begin{equation}
\sqrt{{\rm Ber}^{(2)}_{2N|1}(z;-\kappa)}=\frac{\Delta_{2N}(z)}{\prod_{j=1}^N(z_j+\kappa)(z_j^*+\kappa)}=-\det\left[\begin{array}{c|c} z_a^{b-1} & \displaystyle\underset{}{\frac{1}{z_a+\kappa}} \\  (z_a^*)^{b-1} & \displaystyle\frac{1}{z_a^*+\kappa}  \end{array}\right]_{\substack{1\leq a\leq N\\1\leq b\leq 2N-1}},
\end{equation}
which is the mixture of a Cauchy determinant and a Vandermonde determinant, see Ref.~\onlinecite{BasorForrester1994}. The notation with the vertical line highlights the last column, which consists of rational functions, while the rows have to be understood in pairs, meaning the odd rows consist of $(z_a^0,\ldots,z_a^{2N-2},1/(z_a+\kappa))$ and the even ones are $((z^*_a)^0,\ldots,(z^*_a)^{2N-2},1/(z^*_a+\kappa))$.

It is this determinantal form of the Berezinian, which is useful as we can expand it in the very last column. Due to the permutation symmetry of the integrand in the integration variables $z_j$ as well as their conjugates, each expansion term yields the very same contribution and, hence, an overall factor $2N$ so that we can also write
\begin{equation}\label{Xi2}
\begin{split}
\Xi_2 =&\frac{-2 }{(2\pi)^{N}(N-1)!\prod_{j=1}^{N}{\rm B}(2j,2N+2-2j)}\left(\frac{a_1^2+b_1^2}{b_1a_2-a_1b_2}\right)^{2N}\\
&\times \int\limits_{\mathbb{C}^{N}} d[z] \Delta_{2N-2}(z_1,z_1^*,\ldots,z_{N-1},z_{N-1}^*) \prod_{r=1}^{N} \frac{z_r-z_r^*}{(1+|z_r|^2)^{2N+2}}\ \frac{\prod_{j=1}^{N-1}(z_j-z_N)(z_j^*-z_N)}{z_N^*+\kappa}\\
=&-\frac{2N(2N+1) }{\pi}\left(\frac{a_1^2+b_1^2}{b_1a_2-a_1b_2}\right)^{2N}\int\limits_{\mathbb{C}}d[z_N]\frac{z_N-z_N^*}{(1+|z_N|^2)^{2N+2}}\frac{q_{2N-2}^{(N)}(z_N)}{z_N^*+\kappa}.
\end{split}
\end{equation}
In the second equality we have identified the integral over $z_1,\ldots,z_{N-1}$ with the Heine-formula~\eqref{Heinne.beta4} for $q_{2N-2}^{(N)}(z_N)$.

In expression~\eqref{Xi2} it becomes immediate why the partition function cannot be holomorphic in $a(q_m)$ and $b(q_m)$ anywhere in the complex plane. One can apply the standard formula for the derivative in the complex conjugate $\kappa^*$ on the integral
\begin{equation}
\partial_{\kappa^*}\int\limits_{\mathbb{C}}d[z]\frac{f(z,z^*)}{z+\kappa}\propto f(-\kappa,-\kappa^*)
\end{equation}
for an arbitrary suitably integrable complex function $f(z,z^*)$. Considering the integrand in~\eqref{Xi2} we notice that apart from the real line  the integral must be a function of both, $\kappa$ and $\kappa^*$, which is also what we find. Thus, the analyticity of the integral in $\kappa$ is violated everywhere.

With the help of a similar argument, the remaining integral can be carried out, namely by noticing 
\begin{equation}
\begin{split}
%\frac{\partial}{\partial z_N}\frac{z_N^{2N+1}(z_N^*-z_N)+(1+z_{N}^2)q_{2N}(z_N)}{(1+|z_N|^2)^{2N+1}}=2N(2N+1)\frac{(z_N-z_N^*)q_{2N-2}^{(N)}(z_N)}{(1+|z_N|^2)^{2N+2}}\\
\frac{\partial}{\partial z_N}\frac{z_N^{2N+1}z_N^*+(2N+1)q_{2N}^{(N+1)}(z_N)}{(1+|z_N|^2)^{2N+1}}=2N(2N+1)\frac{(z_N-z_N^*)q_{2N-2}^{(N)}(z_N)}{(1+|z_N|^2)^{2N+2}}
\end{split}
\end{equation}
as well as exploiting the following identity, which is a consequence of the generalized Stokes' theorem (Dolbeault–Grothendieck lemma in Complex Analysis),
\begin{equation}\label{2D-res}
\begin{split}
\int\limits_{\mathbb{C}}d[z_N]\frac{\partial}{\partial z_N}\frac{f(z_N,z_N^*)}{\prod_{j=1}^L(z_N^*+\kappa_j)}=-\pi \sum_{l=1}^L \frac{f(-\kappa_l^*,-\kappa_l)}{\prod_{j\neq l}^L(\kappa_j-\kappa_l)}
\end{split}
\end{equation}
for $L$ distinct $\kappa_j\in\mathbb{C}$ and any differentiable measurable function $f(z_1,z_2)$, which vanishes at infinity in both arguments and where $f(z,z^*)$ is singularity free.
Collecting everything, we find for the function
\begin{equation}
\begin{split}
\Xi_2 =&\left(\frac{a_1^2+b_1^2}{b_1a_2-a_1b_2}\right)^{2N}\biggl[\frac{(\kappa^*)^{2N+1}\kappa+(2N+1)q_{2N}^{(N+1)}(\kappa^*)}{(1+|\kappa|^2)^{2N+1}}\biggl].
\end{split}
\end{equation}
with 
\begin{equation}
\kappa=\frac{a_1a_2+b_1b_2}{b_1a_2-a_1b_2}\quad{\rm and}\quad\kappa^*=\frac{a_1a_2^*+b_1b_2^*}{b_1a_2^*-a_1b_2^*},
\end{equation}
where we have employed the fact that $a_1,b_1\in\mathbb{R}$ are real while $a_2,b_2\in\mathbb{C}$ are complex. The point about which parameter is real or complex is crucial when reinserting the complex functions $(a_1,b_1,a_2,b_2)\to(a(p_n),b(p_n),a(q_m),b(q_m))$ because only $a(q_m)$ and $b(q_m)$ can be complex conjugated while $a(p_n)$ and $b(p_n)$ are analytic continuations of $a_1$ and $b_1$. Therefore, the second kernel is equal to
\begin{equation}\label{K2.result.beta4}
\begin{split}
K_2^{(4)}(p_n,q_m)=&\frac{Z^{(4,N)}_{1|1}(p_n,q_m)}{\kappa(q_m)-\kappa(p_n)}\\
=&\frac{b(p_n)b(q_m)}{a(q_m)b(p_n)-b(q_m)a(p_n)}\left(\frac{a^2(p_n)+b^2(p_n)}{b(p_n)a(q_m)-a(p_n)b(q_m)}\right)^{2N}\\
&\times\frac{\widehat{\kappa}_*^{2N+1}(p_n,q_m)\widehat{\kappa}(p_n,q_m)+(2N+1)q_{2N}^{(N+1)}(\widehat{\kappa}_*(p_n,q_m))}{(1+\widehat{\kappa}(p_n,q_m)\widehat{\kappa}_*(p_n,q_m))^{2N+1}}
\end{split}
\end{equation}
with
\begin{equation}
\widehat{\kappa}(p_n,q_m)=\frac{a(p_n)a(q_m)+b(p_n)b(q_m)}{b(p_n)a(q_m)-a(p_n)b(q_m)}\quad{\rm and}\quad\widehat{\kappa}_*(p_n,q_m)=\frac{a(p_n)a^*(q_m)+b(p_n)b^*(q_m)}{b(p_n)a^*(q_m)-a(p_n)b^*(q_m)}.
\end{equation}
We would like to underline that $\widehat{\kappa}_*(p_n,q_m)$ is not the complex conjugate of $\widehat{\kappa}(p_n,q_m)$, in spite of how we have obtained the expression. It is not immediate from expression~\eqref{K2.result.beta4} that the partition function $Z^{(4,N)}_{1|1}(p_n,q_m)$ is a polynomial in $a(p_n)$ and $b(p_n)$. We only know this from the starting expression in terms of averages over a ratio of two characteristic polynomials of the random matrix $Y$. Anew one can check the ${\rm SO}(2)$ invariance for $Z^{(4,N)}_{1|1}(p_n,q_m)$ which indeed only depends on the group invariants $v^T(p_n)v(p_n)$, $v^T(q_m)v(p_n)$, $v^\dagger(q_m)v(p_n)$, $v^T(q_m)\tau_2v(p_n)$, and $v^\dagger(q_m)\tau_2v(p_n)$.

\subsubsection{The Kernel ${\rm K}_3^{(4)}$}\label{sec:K3.beta4}

For computing the third kernel function, we need to evaluate the integral
\begin{equation}
\begin{split}
\Xi_3=& \frac{1}{(2\pi)^{N}(N+1)!\prod_{j=1}^{N}{\rm B}(2j,2N+2-2j)}\\
&\times\int\limits_{\mathbb{C}^{N+1}} d[z] \Delta_{2N+2}(z) \prod_{r=1}^{N+1} \frac{z_r-z_r^*}{(1+|z_r|^2)^{2N+2}}\ \prod_{j=1}^{N+1}\frac{1}{(\kappa_1 + z_j)(\kappa_1+ z_j^*)(\kappa_2 + z_j)(\kappa_2+ z_j^*)}
\end{split}
\end{equation}
with two distinct complex numbers $\kappa_1,\kappa_2\in\mathbb{C}$. The Vandermonde determinant and the product involving the $\kappa_j$ times the difference $\kappa_2-\kappa_1$ can be written in terms of a  Berezinian, see Ref.~\onlinecite{KieburgGuhr2010a}
\begin{equation}
\begin{split}
\sqrt{{\rm Ber}^{(2)}_{2N+2|2}(z;-\kappa)}=&-\frac{(\kappa_2-\kappa_1)\Delta_{2N+2}(z)}{\prod_{j=1}^{N+1}(\kappa_1 + z_j)(\kappa_1+ z_j^*)(\kappa_2 + z_j)(\kappa_2+ z_j^*)}\\
=&-\det\left[\begin{array}{c|c|c} z_a^{b-1} & \displaystyle\underset{}{\frac{1}{z_a+\kappa_1}}& \displaystyle\underset{}{\frac{1}{z_a+\kappa_2}} \\  (z_a^*)^{b-1} & \displaystyle\frac{1}{z_a^*+\kappa_1}  & \displaystyle\frac{1}{z_a^*+\kappa_2}  \end{array}\right]_{\substack{1\leq a\leq N+1\\1\leq b\leq 2N}}.
\end{split}
\end{equation}
As before the vertical lines should highlight the two last columns, while the odd rows only comprise $z_a$ and the even rows $z_a^*$.
We may choose the skew-orthogonal polynomials $q_j(x)$ in the entries of this determinant instead of the monomials,
\begin{equation}
\begin{split}
\det\left[\begin{array}{c|c|c} z_a^{b-1} & \displaystyle\underset{}{\frac{1}{z_a+\kappa_1}}& \displaystyle\underset{}{\frac{1}{z_a+\kappa_2}} \\  (z_a^*)^{b-1} & \displaystyle\frac{1}{z_a^*+\kappa_1}  & \displaystyle\frac{1}{z_a^*+\kappa_2}  \end{array}\right]_{\substack{1\leq a\leq N+1\\1\leq b\leq 2N}}=\det\left[\begin{array}{c|c|c} q^{(N)}_{b-1}(z_a) & \displaystyle\underset{}{\frac{1}{z_a+\kappa_1}}& \displaystyle\underset{}{\frac{1}{z_a+\kappa_2}} \\  q^{(N)}_{b-1}(z_a^*) & \displaystyle\frac{1}{z_a^*+\kappa_1}  & \displaystyle\frac{1}{z_a^*+\kappa_2}  \end{array}\right]_{\substack{1\leq a\leq N+1\\1\leq b\leq 2N}}.
\end{split}
\end{equation}
This allows us to apply the generalized de Bruijn theorem to carry out the integral, see Ref.~\onlinecite{KieburgGuhr2010a}, yielding
\begin{equation}\label{ber.pol}
\begin{split}
\Xi_3=& \frac{2}{(\kappa_1-\kappa_2)\pi^{N}\prod_{j=1}^{N}{\rm B}(2j,2N+2-2j)}\\
&\times\Pf\left[\begin{array}{c|c|c} \langle q^{(N)}_{a-1}|q^{(N)}_{b-1}\rangle & \displaystyle\left\langle q^{(N)}_{a-1}\left|\frac{1}{z+\kappa_1}\right.\right\rangle & \displaystyle\left\langle q^{(N)}_{a-1}\left|\frac{1}{z+\kappa_2}\right.\right\rangle \\ \hline \displaystyle\left\langle \left. \frac{1}{z+\kappa_1}\right| q^{(N)}_{b-1}\right\rangle & 0 & \displaystyle\left\langle \left. \frac{1}{z+\kappa_1}\right|\frac{1}{z+\kappa_2}\right\rangle \\ \hline \displaystyle\left\langle \left. \frac{1}{z+\kappa_2}\right| q^{(N)}_{b-1}\right\rangle & \displaystyle\left\langle \left. \frac{1}{z+\kappa_2}\right|\frac{1}{z+\kappa_1}\right\rangle & 0 \end{array}\right]_{1\leq a,b\leq 2N}
\end{split}
\end{equation}
where we have employed the skew-symmetric product
\begin{equation}\label{skew-prod}
\braket{f_1|f_2} = \int\limits_{\mathbb{C}} d[z] f_1(z) f_2(z^*) g^{(4)}(z)= -\int\limits_{\mathbb{C}} d[z] f_1(z^*) f_2(z) g^{(4)}(z) = -\braket{f_2|f_1}
\end{equation}
with the weight function
\begin{equation}
g^{(4)}(z)=\frac{z-z^*}{(1+|z|^2)^{2N+2}}.
\end{equation}
This time the vertical and horizontal lines in Eq.~\eqref{ber.pol} emphasize the last two rows and columns. The index $a$ is the row index for the first $2N$ rows and $b$ the column index for the first $2N$ columns. The skew-orthogonality of the polynomials simplifies the upper left $2N\times2N$ block drastically, which becomes a $2\times2$  block-diagonal matrix.
This can be exploited in combination with the standard identity
\begin{equation}
\Pf\left[\begin{array}{cc} A & B \\ -B^T & C \end{array}\right]=\Pf[A]\Pf[C+B^TA^{-1}B]
\end{equation}
to simplify the expression to
\begin{equation}
\begin{split}
(\kappa_2-\kappa_1)\Xi_3=& -2\int\limits_{\mathbb{C}} d[z] \frac{z-z^*}{(1+|z|^2)^{2N+2}} \frac{1}{(\kappa_1 + z)(\kappa_2+ z^*)}\\
&+2\sum_{j=0}^{N-1}\frac{1}{h_j}\left[\left\langle q^{(N)}_{2j}\left|\frac{1}{z+\kappa_1}\right.\right\rangle\left\langle q^{(N)}_{2j+1}\left|\frac{1}{z+\kappa_2}\right.\right\rangle-\left\langle q^{(N)}_{2j+1}\left|\frac{1}{z+\kappa_1}\right.\right\rangle\left\langle q^{(N)}_{2j}\left|\frac{1}{z+\kappa_2}\right.\right\rangle\right]
\end{split}
\end{equation}
with $h_j=1/[\pi{\rm B}(2j+2,2N-2j)]$ being the normalization of the skew-orthogonal polynomials. Plugging in the explicit expressions of the skew-symmetric product and the skew-orthogonal polynomials, we have
\begin{equation}\label{Xi3.int}
\begin{split}
(\kappa_1-\kappa_2)\Xi_3=& -2\int\limits_{\mathbb{C}} d[z] \frac{z-z^*}{(1+|z|^2)^{2N+2}} \frac{1}{(\kappa_1 + z)(\kappa_2+ z^*)}+\frac{2}{\pi}\int\limits_{\mathbb{C}^2}  \frac{d[z_1,z_2](z_1-z_1^*)(z_2-z_2^*)}{(1+|z_1|^2)^{2N+2}(1+|z_2|^2)^{2N+2}} \\
&\times\sum_{j=0}^{N-1}\sum_{m=0}^j\frac{(2N+1)!j!\Gamma(N-j+1/2)}{(2j+1)!(2N-2j-1)!m!\Gamma(N-m+1/2)}\frac{z_1^{2m}z_2^{2j+1}-z_2^{2m}z_1^{2j+1}}{(z_1^*+\kappa_1)(z_2^*+\kappa_2)}.
\end{split}
\end{equation}
In Appendix~\ref{sec:appB}, we evaluate the complex integrals and find
\begin{equation}\label{Xi3}
\begin{split}
(\kappa_1-\kappa_2)\Xi_3=&2\pi(\kappa_1-\kappa_2)\left[\frac{1+\kappa_1^*\kappa_2^*}{(1+|\kappa_1|^2)(1+|\kappa_2|^2)}\right]^{2N+2}\Phi_{2N+2}^{(1)}\left(\frac{|1+\kappa_1\kappa_2|^2}{(1+|\kappa_1|^2)(1+|\kappa_2|^2)}\right)\\
&-2\pi\left(\frac{\kappa_2^*-\kappa_1^*}{(1+|\kappa_1|^2)(1+|\kappa_2|^2)}\right)^{2N+1}q_{2N}^{(N+1)}\left(\frac{\kappa_1^*\kappa_2^*+1}{\kappa_2^*-\kappa_1^*}\right).
%&=-2\pi\left(\frac{\kappa_2^*-\kappa_1^*}{(1+|\kappa_1|^2)(1+|\kappa_2|^2)}\right)^{2N+1}q_{2N}^{(N+1)}\left(\frac{\kappa_1^*\kappa_2^*+1}{\kappa_2^*-\kappa_1^*}\right)-2\pi\frac{\kappa_1-\kappa_2}{(1+\kappa_1\kappa_2)^{2N+2}}\\
%&\times\left[\ln\left(\frac{|\kappa_1-\kappa_2^*|^2}{(1+|\kappa_1|^2)(1+|\kappa_2|^2)}\right)+\sum_{j=1}^{2N+1}\frac{1}{j}\left(\frac{|1+\kappa_1\kappa_2|^2}{(1+|\kappa_1|^2)(1+|\kappa_2|^2)}\right)^j\right]
\end{split}
\end{equation}
$\Phi_{2N+2}^{(1)}(z)$ is Lerch's trancendent~\eqref{Lerch}. Exploiting this result, the third kernel function has the form
\begin{equation}
\begin{split}
{\rm K}_3^{(4)}(q_m,q_n)=&2\pi b(q_m)b(q_n)\biggl[ \left(\frac{b^*(q_m)a^*(q_n)-a^*(q_m)b^*(q_n)}{(|a(q_m)|^2+|b(q_m)|^2)(|a(q_n)|^2+|b(q_n)|^2)}\right)^{2N+1}\\
&\hspace*{-1.6cm}\times q_{2N}^{(N+1)}\left(\frac{a^*(q_m)a^*(q_n)+b^*(q_m)b^*(q_n)}{b^*(q_m)a^*(q_n)-a^*(q_m)b^*(q_n)}\right) - \left[\frac{a^*(q_m)a^*(q_n)+b^*(q_m)b^*(q_n)}{(|a(q_m)|^2+|b(q_m)|^2)(|a(q_n)|^2+|b(q_n)|^2)}\right]^{2N+2}\\
&\hspace*{-1.6cm}\times(b(q_n)a(q_m)-a(q_n)b(q_m))\Phi_{2N+2}^{(1)}\left(\frac{|a(q_m)a(q_n)+b(q_m)b(q_n)|^2}{(|a(q_m)|^2+|b(q_m)|^2)(|a(q_n)|^2+|b(q_n)|^2)}\right)\biggl].
\end{split}
\end{equation}
We rewrote this expression in terms of the vector $v(p)$ in Sec.~\ref{secIII} to underline the invariance under ${\rm SO}(2)$ transformations.

\section{Conclusions}
\label{secV}

We studied statistical aspects of the winding number, which is a
fundamental topological invariant for chiral Hamilton operators. To do so, we set up schematic
models involving two matrices with chiral
unitary (AIII) and symplectic (CII) symmetry and one-dimensional parametric dependence. In particular,
ensemble averages for ratios of determinants with parametric
dependence were computed and related to the $k$-point correlators of the winding number densities. We mapped this problem to averages for ratios of
characteristic polynomials for the respective spherical ensembles and employed
techniques from orthogonal and skew-orthogonal polynomial theory. We verified our analytical results carefully with numerical calculations. We are certain that similar techniques may help to unravel the technically more involved chiral orthogonal symmetry class (BDI). One problem that needs to be addressed in this class is the splitting of the eigenvalues into real and complex conjugate pairs. The $k$-point correlation functions of the corresponding spherical ensemble have been already computed in Refs.~\onlinecite{Krishnapur2009,Mays2013}.

In a previous work~\cite{BHWGG2021}, we also addressed the important
issue of universality, suggesting for the chiral unitary case that the
two-point correlator of the winding number density and the
winding number distribution are universal on proper scales when taking
the limit of infinite matrix dimension. Universality is the crucial
feature making Random Matrix Theory so powerful, see Refs.~\onlinecite{GuhrRMTReview,
  MehtaBook}. Consequently, universality is also a crucial issue in
the new context of statistics for winding numbers and other
topological quantities. At least two questions become relevant: First,
which probability densities of the random matrices are compatible with
universal results and, second, which realizations of the parametric
dependence are admissible? Thorough investigation is beyond the scope of the present
contribution. This includes the rather involved evaluation of all $k$-point correlators for the
winding number density by calculating the proper derivatives
of the formulae we obtained here. In addition, the large $N$-limit
in all possible double scaling limits has to be performed. In a future work we want to address this in combination with universality studies.

Related to the analyzing universality is the following observation.
Our method to explicitly calculate the ensemble averages would also
work for other joint probability density functions of the eigenvalues,
provided the underlying symmetries are the same. This is a
considerable advantage when tackling the problem of universality. In
the ``true'' supersymmetry method that actually employs superspace,
non--Gaussian probability densities for the random matrices can be
treated,
too~\cite{Guhr2006,LittelmannSommersZirnbauer2008,KieburgGronqvistGuhr2009,KieburgSommersGuhr2009}.
Nevertheless, the resulting formulae are less explicit. It is tempting
to speculate that studies along the lines just sketched might help to
improve these results for the ``true'' supersymmetry method.

\begin{acknowledgments}
  We thank Boris Gutkin for fruitful discussions.  This work was
  funded by the German--Israeli Foundation within the project
  \textit{Statistical Topology of Complex Quantum Systems}, grant
  number GIF I-1499-303.7/2019 (N.H.,O.G. and T.G.). Furthermore, M.K. acknowledges support by the Australian
Research Council via Discovery Project grant DP210102887.
\end{acknowledgments}

\appendix

\section{Skew-orthogonal polynomials of the case CII}
\label{AppendixA}

The skew-orthogonal polynomials $q_n^{(N)}$  are defined by choosing them of degree $n$ and the relations
\begin{align}
\braket{q_{2j}^{(N)}|q_{2l}^{(N)}} = \braket{q_{2j+1}^{(N)}|q_{2l+1}^{(N)}} = 0,
\qquad
\braket{q_{2j}^{(N)}|q_{2l+1}^{(N)}} = h_j^{(N)} \delta_{jl}\qquad{\rm for\ all}\ j,l=0,\ldots,N-1,
\end{align}
where we employ the skew-symmetric product~\eqref{skew-prod}.
The normalization constants
\begin{equation}
h_n^{(N)}=\pi\ {\rm B}(2n+2,2N-2n)
\end{equation}
 are related to the normalization $c^{(4)}$ of the joint probability density~\eqref{sphericalev.beta4} in the standard way, see Ref.~\onlinecite{MehtaBook,AKP2010}, namely by
\begin{equation}
c^{(4)}=2^NN!\prod_{j=0}^{N-1}h_j^{(N)}.
\end{equation}
It is well-known, see Refs.~\onlinecite{MehtaBook,AKP2010}, that there is some kind of gauging possible for the polynomials of odd degree by adding a multiple of the even ones ($q_{2j+1}^{(N)}(z)\to q_{2j+1}^{(N)}(z)+c_jq_{2j}^{(N)}(z)$ for any $c_j\in\mathbb{C}$) without destroying the skew-orthogonality. This creates an ambiguity even when choosing  monic normalization  $q_j^{(N)}(x)=x^j+\ldots$ like we will do.

This kind of ambiguity can be fixed by choosing the Heine-like formulae, see Ref.~\onlinecite{AKP2010}, for these polynomials, which are
\begin{align}
q_{2n}^{(N)}(x) &= \frac{\int\limits_{\mathbb{C}^n} d[z] \Delta_{2n}(z) \prod_{j=1}^n  g^{(4)}(z_j)\ \prod_{j=1}^{n} (z_j-x) (z_j^*-x)}{\int\limits_{\mathbb{C}^n} d[z] \Delta_{2n}(z) \prod_{m=1}^n  g^{(4)}(z_m)},\label{Heinne.beta4}
\\
q_{2n+1}^{(N)}(x) &= \frac{\int\limits_{\mathbb{C}^n} d[z] \Delta_{2n}(z) \prod_{j=1}^n  g^{(4)}(z_j)\ \left(x+\sum_{j=1}^{n} [z_j+z_j^*]\right)\prod_{j=1}^{n} (z_j-x) (z_j^*-x)}{\int\limits_{\mathbb{C}^n} d[z] \Delta_{2n}(z) \prod_{m=1}^n  g^{(4)}(z_m)}.
\end{align}
The skew-orthogonal polynomials of even degree are evaluated as follows
\begin{align}	\label{AEvenSkewOrthoPoly}
q_{2n}^{(N)}(x) &\propto \int\limits_{\mathbb{C}^n} d[z] \Delta_{2n+1}(x,z,z^*) \prod_{j=1}^n  \frac{z_j-z_j^*}{\left(1+\abs{z_j}^2\right)^{2N+2}}\propto \Pf \left[\begin{array}{c|c}
0 & x^{b-1}
\\ \hline
-x^{a-1} & D_{ab}
\end{array}\right]_{\substack{1\leq a\leq 2n+1
\\ 1\leq b\leq 2n+1}},
\end{align}
where we have  employed the generalized form of de Bruijn's theorem, see Refs.~\onlinecite{deBruijn1955,KieburgGuhr2010a}, in the second expression and dropped the normalization, which can be reintroduced at the end by employing the monic normalization. The vertical and horizontal line underline the first row and column and $a$ is the running index for the last $2n+1$ rows and $b$ those of the columns. The Pfaffian involves an antisymmetric $(2n+1)\times (2n+1)$-kernel with the elements
\begin{equation}	\label{ADKernel}
\begin{split}
D_{ab} =& 2\int\limits_{\mathbb{C}} d[z] \frac{(z-z^*) z^{a-1} (z^*)^{b-1} 
}{\left(1+\abs{z}^2\right)^{2N+2}}=2 \pi B\left( 2N+2 - \frac{a+b+1}{2},\frac{a+b+1}{2} \right) \left( \delta_{a,b-1}-\delta_{a-1,b}  \right).
\end{split}
\end{equation}
After expanding the Pfaffian in the last row and column we obtain a recursion relation
\begin{equation}
\begin{split}
\Pf \left[\begin{array}{c|c}
0 & x^{b-1}
\\ \hline
-x^{a-1} & D_{ab}
\end{array}\right]_{\substack{1\leq a\leq 2n+1
\\ 1\leq b\leq 2n+1}} =& -\Pf \left[D_{ab}\right]_{\substack{1\leq a,b\leq 2n}}\ x^{2n} +D_{2n,2n+1} \Pf \left[\begin{array}{c|c}
0 & x^{b-1}
\\ \hline
-x^{a-1} & D_{ab}
\end{array}\right]_{\substack{1\leq a\leq 2n-1
\\ 1\leq b\leq 2n-1}}\\
=&-(2\pi)^n \sum_{m=0}^n \prod_{j=1}^{m} B(2N+2-2j,2j) \prod_{j=m+1}^{n} B(2N-2j+1,2j+1) x^{2m}\\
=&-(2\pi)^n\prod_{j=1}^{n} B(2N+2-2j,2j)\ \sum_{m=0}^n\frac{{\rm B}(n+1,N-n+1/2)}{{\rm B}(m+1,N-m+1/2)}x^{2m},
\end{split}
\end{equation}
where we have used
\begin{equation}
\Pf \left[D_{ab}\right]_{\substack{1\leq a,b\leq 2n}}=\prod_{j=0}^{n-1} h_j^{(N)} =(2\pi)^n\prod_{j=1}^n {\rm B}(2j,2N+2-2j)
\end{equation}
After proper normalization we find Eq.~\eqref{skewortho.beta4}.

The calculation of the skew-orthogonal polynomials of odd degree works along the same lines with the only difference of the need for the identity
\begin{align}
\Delta_{2n+1}(x,z,z^*) \left( x + \sum_{j=1}^n [z_j + z_j^*] \right) = \det\left[\begin{array}{c|c}
 \underset{}{z_a^{b-1}} & z_a^{2n+1}  \\
  \overset{}{\underset{}{(z_a^*)^{b-1}}}  &  (z_a^*)^{2n+1}  \\ \hline
 \overset{}{x^{b-1}}  & x^{2n+1}
\end{array}\right]_{\substack{1\leq a\leq n \\ 1\leq b\leq 2n}},
\end{align}
where the vertical and horizontal line highlights the last column and row and the first $n$ odd and even rows comprise $z_a$ and $z_a^*$, respectively.
The polynomials of odd degree are then 
\begin{equation}
\begin{split}
q_{2n+1}^{(N)}(x) \propto& \int\limits_{\mathbb{C}^n} d[z] \Delta_{2n+1}(x,z,z^*) \left(x+\sum_{j=1}^n [z_j + z_j^*]\right) \prod_{j=1}^n  \frac{z_j-z_j^*}{\left(1+\abs{z_j}^2\right)^{2N+2}}\\
\propto&\Pf \left[\begin{array}{c|c|c}
0 & x^{b-1} & x^{2n+1}
\\ \hline
\overset{}{-x^{a-1}} & D_{ab} & 0
\\ \hline
\overset{}{-x^{2n+1}} & 0 & 0
\end{array}\right]_{\substack{1\leq a\leq 2n\\1\leq b \leq 2n}},
\end{split}
\end{equation}
where we anew applied the generalized de Bruijn theorem, see Refs.~\onlinecite{deBruijn1955,KieburgGuhr2010a}. This time the two vertical and horizontal lines underline the particular role of the first and last columns and rows.
The antisymmetric kernel is the same as in the even case \eqref{AEvenSkewOrthoPoly} for $1\leq a,b\leq 2n$. The integrals in the last row and column are the skew-symmetric product $\langle z^{a-1}|z^{2n+1}\rangle$ with $a=1,\ldots,2n$ and, thus, vanish. Expanding the Pfaffian in the last row and column yields the monomial
\begin{align}
q_{2n+1}^{(N)}(x) =x^{2n+1}.
\end{align}
These skew-orthogonal polynomials have to be seen in contrast to those derived in Ref.~\onlinecite{Mays2013} where the author has first mapped the spherical ensemble to a different matrix ensemble. This is the reason why the author of Ref.~\onlinecite{Mays2013} has found the monomials also for the polynomials of even degree.

\section{Evaluating $\Xi_3$}\label{sec:appB}

To simplify expression~\eqref{Xi3.int}, we pursue the same ideas as for the second kernel function. One can show
\begin{equation}
\begin{split}
&\frac{\partial^2}{\partial z_1\partial z_2}\sum_{j=0}^{N}\sum_{m=0}^j\frac{(2N+1)!j!\Gamma(N-j+3/2)}{(2j+1)!(2N-2j+1)!m!\Gamma(N-m+3/2)}\frac{z_1^{2m}z_2^{2j+1}-z_2^{2m}z_1^{2j+1}}{(1+|z_1|^2)^{2N+1}(1+|z_2|^2)^{2N+1}}\\
=&(z_1-z_1^*)(z_2-z_2^*)\sum_{j=0}^{N-1}\sum_{m=0}^j\frac{(2N+1)!j!\Gamma(N-j+1/2)}{(2j+1)!(2N-2j-1)!m!\Gamma(N-m+1/2)}\frac{z_1^{2m}z_2^{2j+1}-z_2^{2m}z_1^{2j+1}}{(1+|z_1|^2)^{2N+2}(1+|z_2|^2)^{2N+2}}\\
&+(2N+1)\frac{(z_2^*-z_1^*)(1+z_1z_2)^{2N}}{(1+|z_1|^2)^{2N+2}(1+|z_2|^2)^{2N+2}}.
\end{split}
\end{equation}
This derivative can be found by recognizing
\begin{equation}
\begin{split}
(1+|z_1|^2)^{2N+2}\frac{\partial}{\partial z_1}\frac{z_1^{2m}}{(1+|z_1|^2)^{2N+1}}=&2m z_1^{2m-1}-(2N-2m+1)z_1^*z_1^{2m},\\
(1+|z_2|^2)^{2N+2}\frac{\partial}{\partial z_2}\frac{z_2^{2j+1}}{(1+|z_2|^2)^{2N+1}}=&(2j+1) z_2^{2j}-(2N-2j)z_2^*z_2^{2j+1}
\end{split}
\end{equation}
which leads to telescopic sums when taking the difference of the left hand side and the first term on the right hand side.

The very first term is the integrand of the twofold integral apart from the factor $1/[(z_1^*+\kappa_1)(z_2^*+\kappa_2)]$. Making use of identity~\eqref{2D-res} for both integration variables $z_1$ and $z_2$ for the left hand side of the equation above, we find
\begin{equation}
\begin{split}
(\kappa_2-\kappa_1)\Xi_3=& -2\int\limits_{\mathbb{C}} d[z] \frac{z-z^*}{(1+|z|^2)^{2N+2}} \frac{1}{(\kappa_1 + z)(\kappa_2+ z^*)}\\
&-\frac{2(2N+1)}{\pi}\int\limits_{\mathbb{C}^2} d[z_1,z_2] \frac{1}{(z_1^*+\kappa_1)(z_2^*+\kappa_2)}\frac{(z_2^*-z_1^*)(1+z_1z_2)^{2N}}{(1+|z_1|^2)^{2N+2}(1+|z_2|^2)^{2N+2}} \\
&\hspace*{-1.5cm}-2\pi\sum_{j=0}^{N}\sum_{m=0}^j\frac{(2N+1)!j!\Gamma(N-j+3/2)}{(2j+1)!(2N-2j+1)!m!\Gamma(N-m+3/2)}\frac{(\kappa_1^*)^{2m}(\kappa_2^*)^{2j+1}-(\kappa_2^*)^{2m}(\kappa_1^*)^{2j+1}}{(1+|\kappa_1|^2)^{2N+1}(1+|\kappa_2|^2)^{2N+1}}.
\end{split}
\end{equation}
The double sum is, apart from the factor $1/[(1+|\kappa_1|^2)^{2N+1}(1+|\kappa_2|^2)^{2N+1}]$, equivalent to an expectation value,
\begin{equation}
\begin{split}
&\sum_{j=0}^{N}\sum_{m=0}^j\frac{(2N+1)!j!\Gamma(N-j+3/2)}{(2j+1)!(2N-2j+1)!m!\Gamma(N-m+3/2)}\left[(\kappa_1^*)^{2m}(\kappa_2^*)^{2j+1}-(\kappa_2^*)^{2m}(\kappa_1^*)^{2j+1}\right]\\
=&\frac{\kappa_2^*-\kappa_1^*}{(2\pi)^{N}N!\prod_{j=1}^{N}{\rm B}(2j,2N+4-2j)}\\
&\times\int\limits_{\mathbb{C}^N} d[z] \Delta_{2N}(z) \prod_{r=1}^N \frac{z_r-z_r^*}{(1+|z_r|^2)^{2N+4}}\ \prod_{j=1}^N(\kappa_1^* + z_j) (\kappa_1^* + z_j^*)(\kappa_2^* + z_j) (\kappa_2^* + z_j^*)\\
=&(\kappa_2^*-\kappa_1^*)\frac{\left\langle\det(\kappa_1^*K_1+K_2)(\kappa_2^*K_1+K_2)\right\rangle_{2N\times 2N}}{\left\langle\det K_1^2\right\rangle_{2N\times 2N}},
\end{split}
\end{equation}
where the subscript $2N\times 2N$ highlights that we average over $2N\times 2N$ real quaternion Ginibre matrices $K_1,K_2\in{\rm Gl}_{\mathbb{H}}(2N)$. We emphasize that we can exploit the results of the first kernel function $K_1^{(4)}(p_m,p_n)$, see Eq.~\eqref{kernel.beta4.K1}, with the difference that the matrix dimension is larger. Thus, it is
\begin{equation}
\begin{split}
&\sum_{j=0}^{N}\sum_{m=0}^j\frac{(2N+1)!j!\Gamma(N-j+3/2)}{(2j+1)!(2N-2j+1)!m!\Gamma(N-m+3/2)}\left[(\kappa_1^*)^{2m}(\kappa_2^*)^{2j+1}-(\kappa_2^*)^{2m}(\kappa_1^*)^{2j+1}\right]\\
%=&(\kappa_2^*-\kappa_1^*)\sum_{j=0}^{N}\frac{{\rm B}(N+1,3/2)}{{\rm B}(j+1,N-j+3/2)}[\kappa_1^*\kappa_2^*+1]^{2j}[\kappa_2^*-\kappa_1^*]^{2N-2j}\\
%=&\frac{(\kappa_2^*-\kappa_1^*)^{2N+1}}{2N+1}\left[\left(1+\left(\frac{\kappa_1^*\kappa_2^*+1}{\kappa_2^*-\kappa_1^*}\right)^2\right)q_{2N}\left(\frac{\kappa_1^*\kappa_2^*+1}{\kappa_2^*-\kappa_1^*}\right)-\left(\frac{\kappa_1^*\kappa_2^*+1}{\kappa_2^*-\kappa_1^*}\right)^{2n+2}\right]\\
=&(\kappa_2^*-\kappa_1^*)^{2N+1}q_{2N}^{(N+1)}\left(\frac{\kappa_1^*\kappa_2^*+1}{\kappa_2^*-\kappa_1^*}\right).
\end{split}
\end{equation}
In addition, the remaining two-fold integral can be simplified further. For that purpose, we note that
\begin{equation}
\frac{\partial}{\partial z_1}\frac{(z_2^*-z_1^*)(1+z_1z_2)^{2N+1}}{(z_1^*+\kappa_1)(z_2-z_1^*)(1+|z_1|^2)^{2N+1}}=(2N+1)\frac{(z_2^*-z_1^*)(1+z_1z_2)^{2N}}{(z_1^*+\kappa_1)(1+|z_1|^2)^{2N+2}}.
\end{equation}
Therefore, we can also evaluate the respective integral for these derivatives along~\eqref{2D-res} where we need to take into account the two poles at $z_1=-\kappa_1^*$ and $z_1=z_2^*$, such that we arrive at
\begin{equation}
\begin{split}
(\kappa_2-\kappa_1)\Xi_3=&-2\pi\left(\frac{\kappa_2^*-\kappa_1^*}{(1+|\kappa_1|^2)(1+|\kappa_2|^2)}\right)^{2N+1}q_{2N}^{(N+1)}\left(\frac{\kappa_1^*\kappa_2^*+1}{\kappa_2^*-\kappa_1^*}\right)\\
&+ 2\int\limits_{\mathbb{C}} d[z] \frac{1}{(1+|z|^2)^{2N+2}}\frac{z^*+\kappa_1}{(z+\kappa_1)(z^*+\kappa_2)}\left(\frac{1-\kappa_1^*z}{1+|\kappa_1|^2}\right)^{2N+1}.
%&-\frac{2\pi}{2N+1}\left(\frac{\kappa_2^*-\kappa_1^*}{(1+|\kappa_1|^2)(1+|\kappa_2|^2)}\right)^{2N+1}\left[\left(1+\left(\frac{\kappa_1^*\kappa_2^*+1}{\kappa_2^*-\kappa_1^*}\right)^2\right)q_{2N}\left(\frac{\kappa_1^*\kappa_2^*+1}{\kappa_2^*-\kappa_1^*}\right)-\left(\frac{\kappa_1^*\kappa_2^*+1}{\kappa_2^*-\kappa_1^*}\right)^{2n+2}\right]
\end{split}
\end{equation}
Extending $z^*+\kappa_1=z^*+\kappa_2+\kappa_1-\kappa_2$ in the numerator, it is straightforward to show that the integral
\begin{equation}
\int\limits_{\mathbb{C}} d[z] \frac{1}{(1+|z|^2)^{2N+2}}\frac{1}{(z+\kappa_1)}\left(\frac{1-\kappa_1^*z}{1+|\kappa_1|^2}\right)^{2N+1}=0
\end{equation}
vanishes, for instance, with the help of Stokes' theorem with
\begin{equation}
\frac{\partial}{\partial z^*}\frac{(1-\kappa_1^*z)^{2N+1}}{z(z+\kappa_1)(1+|z|^2)^{2N+1}}=-\frac{(2N+1)(1-\kappa_1^*z)^{2N+1}}{(z+\kappa_1)(1+|z|^2)^{2N+2}},
\end{equation}
where the contributions at the poles $z=0$ and $z=-\kappa_1$ cancel each other.

What remains is essentially the integral
\begin{equation}\label{J.def}
J=\int\limits_{\mathbb{C}} d[z] \frac{1}{(1+|z|^2)^{2N+2}}\frac{1}{(z+\kappa_1)(z^*+\kappa_2)}\left(\frac{1-\kappa_1^*z}{1+|\kappa_1|^2}\right)^{2N+1}.
\end{equation}
Choosing polar coordinates $z=\sqrt{r}e^{i\varphi}$, we first integrate over the angle $\varphi\in[0,2\pi]$, exploiting the partial fraction decomposition
\begin{equation}
\frac{1}{(\sqrt{r}e^{i\varphi}+\kappa_1)(\sqrt{r}e^{-i\varphi}+\kappa_2)}=\frac{e^{i\varphi}}{r-\kappa_1\kappa_2}\left[\frac{1}{e^{i\varphi}+\kappa_1/\sqrt{r}}-\frac{1}{e^{i\varphi}+\sqrt{r}/\kappa_2}\right]
\end{equation}
and employing the residue theorem, which leads to
\begin{equation}
J=\pi\int\limits_{|\kappa_1|^2}^\infty \frac{dr}{(1+r)^{2N+2}(r-\kappa_1\kappa_2)}-\pi\int\limits_0^{|\kappa_2|^2} \frac{dr}{(1+r)^{2N+2}(r-\kappa_1\kappa_2)}\left(\frac{1+r\kappa_1^*/\kappa_2}{1+|\kappa_1|^2}\right)^{2N+1}.
\end{equation}
The first integral is explicitly
\begin{equation}
\int\limits_{|\kappa_1|^2}^\infty \frac{dr}{(1+r)^{2N+2}(r-\kappa_1\kappa_2)}=-\frac{1}{(1+\kappa_1\kappa_2)^{2N+2}}\left[\ln\left(1-\frac{1+\kappa_1\kappa_2}{1+|\kappa_1|^2}\right)+\sum_{j=1}^{2N+1}\frac{1}{j}\left(\frac{1+\kappa_1\kappa_2}{1+|\kappa_1|^2}\right)^j\right],
\end{equation}
which is essentially Lerch's transcendent~\eqref{Lerch}.
The second integral can be evaluated once one has performed the M\"obius transformation
\begin{equation}
s=\frac{(\kappa_2-\kappa_1^*)r}{\kappa_2+\kappa_1^*r}\quad\Leftrightarrow\quad r=\frac{\kappa_2 s}{\kappa_2-\kappa_1^*-\kappa_1^*s}.
\end{equation}
Then, the integral simplifies to
\begin{equation}
\begin{split}
&\int\limits_0^{|\kappa_2|^2} \frac{dr}{(1+r)^{2N+2}(r-\kappa_1\kappa_2)}\left(\frac{1+r\kappa_1^*/\kappa_2}{1+|\kappa_1|^2}\right)^{2N+1}\\
=&\int\limits_0^{(|\kappa_2|^2-\kappa_1^*\kappa_2^*)/(1+\kappa_1^*\kappa_2^*)}\frac{ds}{(1+|\kappa_1|^2)^{2N+1}(1+s)^{2N+2}[(1+|\kappa_1|^2)s+|\kappa_1|^2-\kappa_1\kappa_2]}.
\end{split}
\end{equation}
This integral can be carried out like the former one, yielding
\begin{equation}
\begin{split}
&\int\limits_0^{|\kappa_2|^2} \frac{dr}{(1+r)^{2N+2}(r-\kappa_1\kappa_2)}\left(\frac{1+r\kappa_1^*/\kappa_2}{1+|\kappa_1|^2}\right)^{2N+1}\\
=&\frac{1}{(1+\kappa_1\kappa_2)^{2N+2}}\left[\ln\left(1-\frac{1+\kappa_1\kappa_2}{1+|\kappa_1|^2}\right)+\sum_{j=1}^{2N+1}\frac{1}{j}\left(\frac{1+\kappa_1\kappa_2}{1+|\kappa_1|^2}\right)^j\right]\\
&-\frac{1}{(1+\kappa_1\kappa_2)^{2N+2}}\left[\ln\left(1-\frac{|1+\kappa_1\kappa_2|^2}{(1+|\kappa_1|^2)(1+|\kappa_2|^2)}\right)+\sum_{j=1}^{2N+1}\frac{1}{j}\left(\frac{|1+\kappa_1\kappa_2|^2}{(1+|\kappa_1|^2)(1+|\kappa_2|^2)}\right)^j\right].
\end{split}
\end{equation}
As can be seen the first logarithm and sum cancel with the one from the first integral of $J$. Therefore, we arrive at
\begin{equation}
\begin{split}
J=&-\frac{\pi}{(1+\kappa_1\kappa_2)^{2N+2}}\left[\ln\left(1-\frac{|1+\kappa_1\kappa_2|^2}{(1+|\kappa_1|^2)(1+|\kappa_2|^2)}\right)+\sum_{j=1}^{2N+1}\frac{1}{j}\left(\frac{|1+\kappa_1\kappa_2|^2}{(1+|\kappa_1|^2)(1+|\kappa_2|^2)}\right)^j\right]
\\
&= \pi \left( \frac{1+\kappa_1^* \kappa_2^*}{(1+|\kappa_1|^2)(1+|\kappa_2|^2)} \right)^{2N+2} \Phi_{2N+2}^{(1)}\left( \frac{|1+\kappa_1\kappa_2|^2}{(1+|\kappa_1|^2)(1+|\kappa_2|^2)} \right),
\end{split}
\end{equation}
which is anew Lerch's transcendent~\eqref{Lerch} apart from a prefactor.
Despite that some expressions of this integral has been in some intermediate steps not obviously symmetric under $\kappa_1\leftrightarrow\kappa_2$, this final result reflects this symmetry.

\bibliography{stz}

\end{document}